\documentstyle [twoside,12pt]{article}
\begin{document}
\begin{flushright}
SADF1-1995
\end{flushright}
\sloppy
\centerline{{\Large\bf Quantum field theory of fermion mixing}}

\vspace{.4in}
\centerline{{\bf M. Blasone\footnote{blasone@sa.infn.it}
and G. Vitiello\footnote{vitiello@sa.infn.it \\   manuscript of 28 pages}}}

\vspace{.2in}
\centerline{Dipartimento di Fisica dell'Universita'}
\centerline{and INFN, Gruppo Collegato, Salerno}
\centerline{I-84100 Salerno, Italy}

\vspace{.6in}

\centerline{{\bf Abstract}}

\bigskip

\small
The fermion mixing transformations are studied in the quantum field theory
framework. In particular neutrino mixing is considered and the Fock space of
definite flavor states is shown to be unitarily inequivalent to the Fock space
of definite mass states. The flavor oscillation formula is computed for two
and three flavors mixing and the oscillation amplitude is found to be momentum
dependent, a result which may be subject to experimental test. The flavor
vacuum state exhibits the structure of $SU(2)$ generalized coherent state.
\normalsize

\bigskip

P.A.C.S. 11.10.-z , 11.30.Jw , 14.60.Gh

\thispagestyle{empty}

\newpage
{\bf 1  Introduction}

\medskip

Mixing transformations of fermion fields play a crucial r\^ole in high energy
physics. The original Cabibbo mixing of d and s quarks and its extension to
the Kobayashi-Maskawa three flavors mixing are essential ingredients in the
Standard Model phenomenology [1].
On the other hand, although clear experimental
evidence is still missing, it is widely believed that neutrino mixing
transformations are the basic tool for further understanding of neutrino
phenomenology as well as of solar physics [2].

In contrast with the large body of successful modelling and phenomenological
computations (especially for Cabibbo-Kobayashi-Maskawa (CKM)
quark mixing, the neutrino mixing still waiting for
a conclusive experimental evidence), the quantum field
theoretical analysis of the mixing transformations has not been pushed much
deeply, as far as we know. The purpose of the present paper is indeed the study
of the quantum field theory (QFT) framework of the fermion mixing
transformations, thus focusing our attention more on the theoretical
structure of fermion mixing than on its phenomenological features.

As we will see, our study is far from being purely academic since, by
clarifying
the theoretical framework of fermion mixing, we will obtain some results which
are also interesting to phenomenology and therefore to the real life of
experiments.

In particular, to be definite, we will focus our attention
on neutrino mixing transformations and our analysis will lead to some
modifications of the neutrino oscillation formulas, which in fact may be
subject to experimental test.

The paper is organized as follows. In Section 2 we study the generator of the
Pontecorvo neutrino mixing transformations (two flavors mixing for Dirac
fields). We show that in the Lehmann-Symanzik-Zimmermann
(LSZ) formalism of quantum field theory [3,6,7] the Fock
space of the flavor states is unitarily inequivalent to the Fock space of the
mass eigenstates in the infinite volume limit. The flavor states are obtained
as condensate of massive neutrino pairs and exhibit the structure of $SU(2)$
coherent states [4]. In Section 3 we exhibit the condensation
density as a function of the mixing angle, of the momentum
and of the neutrino masses. In Section 4 we derive the neutrino
flavor oscillations whose amplitude turns out to be momentum and mass
dependent.
This is a novel feature with respect to conventional analysis and
may be subject to experimental test. In some sense, from the point of view
of phenomenology, this is the most interesting result. Nevertheless, the
condensate and coherent state structure of the vacuum is by itself a novel
and theoretically interesting feature emerging from our analysis. In Section
5 we extend our considerations to three flavors mixing and show how the
transformation matrix is obtained in terms of the QFT generators introduced in
Section 2. We also obtain the three flavors oscillation formula.
Finally, Section 6 is devoted to the conclusions. Although the group
theoretical analysis is conceptually simple, specific computations are sometime
lengthy and, for the reader convenience, we confine mathematical details to
the Appendices.

\bigskip
\bigskip
{\bf 2  The vacuum structure for fermion mixing}

\medskip

For definitiveness we consider the Pontecorvo mixing relations [5],
although the following discussion applies to any Dirac fields.

The mixing relations are:

$$\nu_{e}(x) = \nu_{1}(x) \; \cos\theta   + \nu_{2}(x) \; \sin\theta  $$
$$\nu_{\mu}(x) =- \nu_{1}(x) \; \sin\theta   + \nu_{2}(x)\;
\cos\theta\;,  \eqno(2.1)$$
where $\nu_{e}(x)$ and $\nu_{\mu}(x)$ are the (Dirac) neutrino fields
with definite flavors.
$\nu_{1}(x)$ and $\nu_{2}(x)$ are the (free) neutrino
fields with definite masses $m_{1}$ and $m_{2}$, respectively.
The fields $\nu_{1}(x)$ and $\nu_{2}(x)$ are written as
$$\nu_{i}(x) = \frac{1}{\sqrt{V}} \sum_{\bar{k},r}[u^{r}_{\bar{k},i}
\alpha ^{r}_{\bar{k},i}\:e^{i \bar{k}\cdot \bar{x}}+ v^{r}_{\bar{k},i}
\beta ^{r\dag }_{\bar{k},i}\: e^{-i \bar{k}\cdot \bar{x}}], \; ~ i=1,2 \;.
\eqno(2.2) $$
In the following, for simplicity, we will omit the vector notation for
$\bar{k}$ and use the same symbol $k$ to denote both $\bar{k}$ and its modulus
$k$.
$\alpha ^{r}_{k,i}$ and $ \beta ^{r }_{k,i}$, $  i=1,2 \;, \;r=1,2$
are the annihilator operators for the vacuum state
$|0\rangle_{1,2}\equiv|0\rangle_{1}\otimes |0\rangle_{2}$:
$\alpha ^{r}_{k,i}|0\rangle_{12}= \beta ^{r }_{k,i}|0\rangle_{12}=0$.
In eq.(2.2) we have included the time dependence in the wave functions. In the
following this dependence will be omitted when no misunderstanding arises.
The anticommutation relations are:
$$\{\nu^{\alpha}_{i}(x), \nu^{\beta\dag }_{j}(y)\}_{t=t'} = \delta^{3}(x-y)
\delta _{\alpha\beta} \delta_{ij} \;, \;\;\;\;\; \alpha,\beta=1,..,4 \;,
\eqno(2.3)$$
and
$$\{\alpha ^{r}_{k,i}, \alpha ^{s\dag }_{q,j}\} = \delta
_{kq}\delta _{rs}\delta _{ij}   ;\qquad \{\beta ^{r}_{k,i}, \beta ^{s\dag
}_{q,j}\} = \delta _{kq} \delta _{rs}\delta _{ij},\;\;\;\;
i,j=1,2\;. \eqno(2.4)  $$
All other anticommutators are zero. The orthonormality and
completeness relations are:
$$ \sum_{\alpha}u^{r\alpha*}_{k,i} u^{s\alpha}_{k,i} = \sum_{\alpha}
v^{r\alpha*}_{k,i} v^{s\alpha}_{k,i} = \delta_{rs}
\;,\qquad  \sum_{\alpha}u^{r\alpha*}_{k,i} v^{s\alpha}_{-k,i} =
\sum_{\alpha}v^{r\alpha*}_{-k,i} u^{s\alpha}_{k,i} = 0\;, $$
\smallskip
$$\sum_{r}(u^{r\alpha*}_{k,i} u^{r\beta}_{k,i} +
v^{r\alpha*}_{-k,i} v^{r\beta}_{-k,i}) = \delta_{\alpha\beta}\;. \eqno(2.5) $$
Eqs.(2.1) (or the ones obtained by inverting them) relate the respective
hamiltonians $H_{1,2}$ (we consider only the mass terms) and $H_{e,\mu}$ [5]:
$$H_{1,2}=m_{1}\;\nu^{\dag}_{1} \nu_{1} + m_{2}\;\nu^{\dag}_{2} \nu_{2}
\eqno(2.6)$$
$$H_{e,\mu}=m_{ee}\;\nu^{\dag}_{e} \nu_{e} +
m_{\mu\mu}\;\nu^{\dag}_{\mu} \nu_{\mu}+
m_{e\mu}\left(\nu^{\dag}_{e} \nu_{\mu} + \nu^{\dag}_{\mu} \nu_{e}\right)
\eqno(2.7)$$
where $m_{ee}=m_{1}\cos^{2}\theta + m_{2} \sin^{2} \theta$,
$m_{\mu\mu}=m_{1}\sin^{2}\theta + m_{2} \cos^{2} \theta$
and $m_{e\mu}=(m_{2}-m_{1})\sin\theta \cos \theta$.

In QFT the basic dynamics, i.e. the Lagrangian and the resulting field
equations,
is given in terms of Heisenberg (or interacting) fields. The physical
observables are expressed in terms of asymptotic in- (or out-) fields, also
called physical or free fields. In the LSZ formalism
of QFT [3,6,7],
the free fields, say for definitiveness the in-fields, are obtained
by the weak limit of the Heisenberg fields for time $t \rightarrow - \infty$.
The meaning of the weak limit is that the realization of the basic dynamics in
terms of the in-fields is not unique so that the limit for $t \rightarrow -
 \infty$ (or $t \rightarrow + \infty$ for the out-fields) is representation
dependent. Typical examples are the ones of spontaneously broken symmetry
theories, where the same set of Heisenberg field equations describes the normal
(symmetric) phase as well as the symmetry broken phase.
The representation dependence
of the asymptotic limit arises from the existence in QFT of infinitely many
unitarily non-equivalent representations of the canonical (anti-)commutation
relations [6,7].
Of course, since observables are described in terms of asymptotic
fields, unitarily inequivalent representations describe different, i.e.
physically inequivalent, phases. It is therefore of crucial importance, in
order to get physically meaningful results, to investigate with much care the
mapping among Heisenberg or interacting fields and free fields.
Such a mapping is usually called the Haag expansion or the dynamical map [6,7].
Only in a very
rude and naive approximation we may assume that interacting fields and free
fields share the same vacuum state and the same Fock space representation.

We stress that the above remarks apply to QFT, namely to systems with infinite
number of degrees of freedom. In quantum mechanics, where finite volume
systems are considered, the von Neumann theorem ensures that the
representations of the canonical commutation relations
are each other unitary equivalent and no problem arises
with uniqueness of the asymptotic limit. In QFT, however, the von Neumann
theorem does not hold and much more careful attention is required when
considering any mapping among interacting and free fields [6,7].

With this warnings, mixing relations such as the relations (2.1) deserve a
careful analysis. It is in fact our purpose to investigate
the structure of the Fock spaces
${\cal H}_{1,2}$ and ${\cal H}_{e,\mu}$ relative to
$\nu_{1}(x)$, $\nu_{2}(x)$ and  $\nu_{e}(x)$,  $\nu_{\mu}(x)$, respectively.
In particular we want to study the relation among these spaces
in the infinite volume limit.
We expect that ${\cal H}_{1,2}$ and ${\cal H}_{e,\mu}$ become
orthogonal in such a limit, since they represent the Hilbert spaces for
free and interacting fields, respectively [6,7].
In the following, as usual, we will perform all computations at finite
volume $V$ and only at the end we will put $V \rightarrow \infty$.

Our first step is the study of the generator of eqs.(2.1) and of the underlying
group theoretical structure.

Eqs.(2.1) can be put in the form:
$$\nu_{e}^{\alpha}(x) = G^{-1}(\theta)\; \nu_{1}^{\alpha}(x)\; G(\theta) $$
$$\nu_{\mu}^{\alpha}(x) = G^{-1}(\theta)\; \nu_{2}^{\alpha}(x)\; G(\theta)
\;, \eqno(2.8)  $$
where $G(\theta)$ is given by
$$ G(\theta) = exp\left[\theta \int d^{3}x  \left(\nu_{1}^{\dag}(x)
\nu_{2}(x) - \nu_{2}^{\dag}(x) \nu_{1}(x)
\right)\right]\;, \eqno(2.9)$$
and is (at finite volume) an unitary operator:
$G^{-1}(\theta)=G(-\theta)=G^{\dag}(\theta)$.
We indeed observe that, from eqs.(2.8),
$d^{2}\nu^{\alpha}_{e}/d\theta^{2}=-\nu^{\alpha}_{e}\;,\;\;\;
d^{2}\nu^{\alpha}_{\mu}/d\theta^{2}=-\nu^{\alpha}_{\mu}$.
By using the initial conditions
$\nu^{\alpha}_{e}|_{\theta=0}=\nu^{\alpha}_{1}$,
$d\nu^{\alpha}_{e}/d\theta|_{\theta=0}=\nu^{\alpha}_{2}$ and
$\nu^{\alpha}_{\mu}|_{\theta=0}=\nu^{\alpha}_{2}$,
$d\nu^{\alpha}_{\mu}/d\theta|_{\theta=0}=-\nu^{\alpha}_{1}$,
we see that $G(\theta)$ generates eqs.(2.1).

By introducing the operators
$$ S_{+} \equiv  \int d^{3}x \; \nu_{1}^{\dag}(x)
\nu_{2}(x) \;\;,\;\;\;
S_{-} \equiv  \int d^{3}x \; \nu_{2}^{\dag}(x)
\nu_{1}(x)\;= \left( S_{+}\right)^{\dag}\;, \eqno(2.10) $$
$G(\theta)$ can be written as
$$
G(\theta) = exp[\theta(S_{+} - S_{-})]\;. \eqno(2.11)$$
It is easy to verify that, introducing $S_{3}$ and the total charge
$S_{0}$ as follows
$$ S_{3} \equiv \frac{1}{2} \int d^{3}x
\left(\nu_{1}^{\dag}(x)\nu_{1}(x) -
\nu_{2}^{\dag}(x)\nu_{2}(x)\right)\;, \eqno(2.12) $$
\smallskip
$$ S_{0} \equiv \frac{1}{2} \int d^{3}x
\left(\nu_{1}^{\dag}(x)\nu_{1}(x) +
\nu_{2}^{\dag}(x)\nu_{2}(x)\right)\;, \eqno(2.13) $$
the $su(2)$ algebra is closed:
$$ [S_{+} , S_{-}]=2S_{3} \;\;\;,\;\;\; [S_{3} , S_{\pm} ] = \pm S_{\pm}
\;\;\;,\;\;\;[S_{0} , S_{3}]= [S_{0} , S_{\pm} ] = 0\;. \eqno(2.14)$$

Using eq.(2.2) we can expand $S_{+}$, $S_{-}$, $S_{3}$ and $S_{0}$ as follows:
$$S_{+}\equiv \sum_{k} S_{+}^{k}=\sum_{k}\sum_{r,s}$$
$$( u^{r\dag}_{k,1} u^{s}_{k,2}\;  \alpha^{r\dag}_{k,1} \alpha^{s}_{k,2}
+ v^{r\dag}_{-k,1} u^{s}_{k,2} \; \beta^{r}_{-k,1}  \alpha^{s}_{k,2}
+ u^{r\dag}_{k,1} v^{s}_{-k,2} \;\alpha^{r\dag}_{k,1}\beta^{s\dag}_{-k,2}
+ v^{r\dag}_{-k,1} v^{s}_{-k,2}\; \beta^{r}_{-k,1}\beta^{s\dag}_{-k,2})\;,
\eqno(2.15)$$
$$S_{-}\equiv \sum_{k} S_{-}^{k}=\sum_{k}\sum_{r,s}$$
$$( u^{r\dag}_{k,2} u^{s}_{k,1}\; \alpha^{r\dag}_{k,2}  \alpha^{s}_{k,1}
+   v^{r\dag}_{-k,2} u^{s}_{k,1} \;\beta^{r}_{-k,2}  \alpha^{s}_{k,1}
+   u^{r\dag}_{k,2} v^{s}_{-k,1} \;\alpha^{r\dag}_{k,2}\beta^{s\dag}_{-k,1}
+   v^{r\dag}_{-k,2} v^{s}_{-k,1}\; \beta^{r}_{-k,2}\beta^{s\dag}_{-k,1})\;,
\eqno(2.16)$$
$$S_{3}\equiv \sum_{k} S_{3}^{k}
=\frac{1}{2}\sum_{k,r}\left(\alpha^{r\dag}_{k,1}\alpha^{r}_{k,1}
-\beta^{r\dag}_{-k,1}\beta^{r}_{-k,1}
-\alpha^{r\dag}_{k,2}\alpha^{r}_{k,2} + \beta^{r\dag}_{-k,2}\beta^{r}_{-k,2}
\right)\;, \eqno(2.17) $$
$$S_{0}\equiv \sum_{k} S_{0}^{k}
=\frac{1}{2}\sum_{k,r}\left(\alpha^{r\dag}_{k,1}\alpha^{r}_{k,1}
-\beta^{r\dag}_{-k,1}\beta^{r}_{-k,1}
+\alpha^{r\dag}_{k,2}\alpha^{r}_{k,2} - \beta^{r\dag}_{-k,2}\beta^{r}_{-k,2}
\right)\;. \eqno(2.18) $$
It is interesting to observe that the operatorial structure of eqs.(2.15) and
(2.16) is the one of the rotation generator and of the Bogoliubov generator.
These structures will be exploited in the following (cf. Section 3 and
Appendix D).
Using these expansions it is easy to show that the following relations hold:
$$ [S_{+}^{k} , S_{-}^{k}]=2S_{3}^{k} \;\;\;,\;\;\;
[S_{3}^{k} , S_{\pm}^{k} ] = \pm S_{\pm}^{k}\;\;\;,
\;\;\;[S_{0}^{k} , S_{3}^{k}]= [S_{0}^{k} , S_{\pm}^{k} ] = 0\;, \eqno(2.19)$$
$$ [S_{\pm}^{k} , S_{\pm}^{p}]= [S_{3}^{k} , S_{\pm}^{p} ] =
[S_{3}^{k} , S_{3}^{p} ] =0 \;\;\;,\;\;\;k \neq p\;. \eqno(2.20)$$
This means that the original $su(2)$ algebra given in eqs.(2.14)
splits into $k$ disjoint $su_{k}(2)$ algebras, given by eqs.(2.19), i.e.
we have the group structure $ \bigotimes_{k} SU_{k}(2)$.

To establish the relation between ${\cal H}_{1,2}$ and ${\cal H}_{e,\mu}$
we consider the generic matrix element
$_{1,2}\langle a|\nu^{\alpha}_{1}(x)|b \rangle_{1,2}$
(a similar argument holds for $\nu^{\alpha}_{2}(x)$),
where $|a \rangle_{1,2}$ is the generic element of ${\cal H}_{1,2}$.
Using the inverse of the first of the (2.8), we obtain:
$$\;_{1,2}\langle a|G(\theta)\; \nu^{\alpha}_{e}(x)\;
G^{-1}(\theta) |b \rangle_{1,2}\;
=\;_{1,2}\langle a|\nu^{\alpha}_{1}(x)|b \rangle_{1,2}\;. \eqno(2.21)$$
Since the operator field $\nu_{e}$ is defined on the Hilbert space
${\cal H}_{e,\mu}$, eq.(2.21) shows that
$G^{-1}(\theta) |a \rangle_{1,2}$ is a vector of
${\cal H}_{e,\mu}$, so $G^{-1}(\theta)$  maps ${\cal H}_{1,2}$
to  ${\cal H}_{e,\mu}$:
$G^{-1}(\theta): {\cal H}_{1,2} \mapsto {\cal H}_{e,\mu}$.
In particular for the vacuum $|0 \rangle_{1,2}$ we have (at finite volume $V$):
$$|0 \rangle_{e,\mu} = G^{-1}(\theta)\; |0 \rangle_{1,2}\;. \eqno(2.22)$$
$|0 \rangle_{e,\mu}$ is the vacuum for ${\cal H}_{e,\mu}$. In fact,
from eqs.(2.8) we obtain the positive frequency operators, i.e.
the annihilators, relative to the fields $\nu_{e}(x)$ and $\nu_{\mu}(x)$ as
$$u^{r \alpha}_{k,e} \;{\tilde\alpha}^{r}_{k,e} =
G^{-1}(\theta)\; u^{r \alpha}_{k,1} \;\alpha^{r}_{k,1}
\;G(\theta)\;,\eqno(2.23a)$$
$$u^{r \alpha}_{k,\mu} \;{\tilde\alpha}^{r}_{k,\mu} =
G^{-1}(\theta)\; u^{r \alpha}_{k,2} \;\alpha^{r}_{k,2}
\;G(\theta)\;,\eqno(2.23b)$$
$$v^{r \alpha *}_{k,e} \;{\tilde\beta}^{r}_{k,e} =
G^{-1}(\theta)\; v^{r \alpha *}_{k,1}
\;\beta^{r}_{k,1}\;G(\theta)\;,\eqno(2.23c)$$
$$v^{r \alpha *}_{k,\mu} \;{\tilde\beta}^{r}_{k,\mu} =
G^{-1}(\theta)\; v^{r \alpha *}_{k,2}
\;\beta^{r}_{k,2}\;G(\theta)\;.\eqno(2.23d)$$
Eqs.(2.23) are obtained by using the linearity of operator $G(\theta)$.
It is a trivial matter to check that these operators do effectively annihilate
$|0\rangle_{e,\mu}$.

Furthermore, for the vacuum state $|0\rangle_{e,\mu}$ the conditions hold:
$$\int d^{3}x
\;\nu_{e}^{\dag}(x)\nu_{e}(x)|0\rangle_{e,\mu}=0\;\;\;,\;\;\;
\int d^{3}x \;\nu_{\mu}^{\dag}(x)\nu_{\mu}(x)|0\rangle_{e,\mu}=0\;,
\eqno(2.24)$$
as can be verified using eqs.(2.8) and (2.22), or the definitions (2.23).

In Section 3 we will
explicitly compute eqs.(2.23), thus giving the dynamical map of the
flavor operators in terms of the mass operators.

We observe that $G^{-1}(\theta) = exp[\theta(S_{-} - S_{+})]$ is just the
generator for generalized coherent states of $SU(2)$: the flavor vacuum
state is therefore an $SU(2)$ coherent state. Let us obtain
the explicit expression for $|0\rangle_{e,\mu}$ and investigate the infinite
volume limit of eq.(2.22).

Using the Gaussian decomposition, $G^{-1}(\theta)$ can be written as [4]
$$exp[\theta(S_{-} - S_{+})]= exp(-tan\theta \; S_{+})
\;exp(-2 ln \: cos\theta \; S_{3})
\;exp(tan\theta \; S_{-}) \eqno(2.25)$$
where $0\leq \theta < \frac{\pi}{2}$. Eq.(2.22) then becomes
$$|0\rangle_{e,\mu} = \prod_{k} exp(-tan\theta \; S_{+}^{k})
exp(-2 ln \: cos\theta \; S_{3}^{k})
\;exp(tan\theta \; S_{-}^{k})|0\rangle_{1,2}\;. \eqno(2.26)$$
The right hand side of eq.(2.26) may be computed by using the relations
$$S_{3}^{k}|0\rangle_{1,2}=0\;\;,\;\;S_{\pm}^{k}|0\rangle_{1,2}\neq 0
\;\;,\;\; (S_{\pm}^{k})^{2}|0\rangle_{1,2}\neq 0 \;\;,\;\;
(S_{\pm}^{k})^{3}|0\rangle_{1,2}=0\;, \eqno(2.27) $$
and other useful relations which are given in the Appendix A.
The final expression for $|0\rangle_{e,\mu}$ in terms of
$S^{k}_{\pm}$ and $S^{k}_{3}$ is:
$$|0\rangle_{e,\mu}=\prod_{k}|0\rangle_{e,\mu}^{k}=
\prod_{k}\left[ 1 + \sin\theta \cos\theta
\left(S_{-}^{k} - S_{+}^{k}\right)+\frac{1}{2}\sin^{2}\theta \cos^{2}\theta
\left((S_{-}^{k})^{2} + (S_{+}^{k})^{2}\right)+\right.$$
$$\left. -\sin^{2}\theta S_{+}^{k}S_{-}^{k} +
\frac{1}{2}\sin^{3}\theta \cos\theta \left(S_{-}^{k}(S_{+}^{k})^{2} -
S_{+}^{k}(S_{-}^{k})^{2}\right)+ \frac{1}{4} \sin^{4}\theta
(S_{+}^{k})^{2}(S_{-}^{k})^{2}\right]|0\rangle_{1,2}\;. \eqno(2.28)$$

The state $|0\rangle_{e,\mu}$ is normalized to 1 (see eq.(2.22)).
Eq.(2.28) and eqs.(2.15) and (2.16) exhibit the rich coherent state structure
of
$|0\rangle_{e,\mu}$.

Let us now compute $_{1,2}\langle0|0\rangle_{e,\mu}$. We obtain
$$_{1,2}\langle0|0\rangle_{e,\mu}  =
\prod_{k}\left(1-\sin^{2}\theta\; _{1,2}\langle0|S_{+}^{k}
S_{-}^{k}|0\rangle_{1,2} + \frac{1}{4}
\sin^{4}\theta\; _{1,2}\langle0|(S_{+}^{k})^{2}
(S_{-}^{k})^{2}|0\rangle_{1,2}\right)\eqno(2.29)$$
where (see Appendix B)
$$_{1,2}\langle0|S_{+}^{k}S_{-}^{k}|0\rangle_{1,2}
=_{1,2}\langle0|\sum_{\sigma,\tau}\sum_{r,s} (v^{\sigma\dag}_{-k,1}
u^{\tau}_{k,2})
(u^{s\dag}_{k,2} v^{r}_{-k,1})
\beta^{\sigma}_{-k,1}\alpha^{\tau}_{k,2}
\alpha^{s\dag}_{k,2}\beta^{r\dag}_{-k,1}|0\rangle_{1,2} = $$
$$=\sum_{r,s} |\;v^{r\dag}_{-k,1} u^{s}_{k,2}\; |^{2} \equiv Z_{k}\;.
\eqno(2.30)$$
\smallskip
In a similar way we find
$$_{1,2}\langle0|(S_{+}^{k})^{2}(S_{-}^{k})^{2}|0\rangle_{1,2}= Z_{k}^{2}\;.
\eqno(2.31)$$
\medskip
Explicitly, $Z_{k}$ is given by
$$Z_{k}=
\frac{k^{2}\left[(\omega_{k,2}+m_{2})-(\omega_{k,1}+m_{1})\right]^{2}}
{2\;\omega_{k,1}\omega_{k,2}(\omega_{k,1}+m_{1})(\omega_{k,2}+m_{2})}\eqno
(2.32)$$
where $\omega_{k,i}=\sqrt{k^{2}+m_{i}^{2}}$.
The function $Z_{k}$ depends on $k$ only through its modulus and it is
always in the interval $[0,1[$. It has a maximum for
$k= \sqrt{m_{1}m_{2}}$ which tends asymptotically to 1 when $ |m_{2} - m_{1}|
\rightarrow \infty$; also, $Z_{k} \rightarrow 0$ when $k \rightarrow \infty$.

In conclusion we have
$$_{1,2}\langle0|0\rangle_{e,\mu}  =\prod_{k}\left(1-
\frac{1}{2} \sin^{2}\theta\;Z_{k}\right)^{2}\equiv
\prod_{k}\Gamma(k)=$$
$$=\prod_{k}e^{ln\; \Gamma(k)}=e^{\sum_{k}ln\; \Gamma(k)}.\eqno(2.33)$$
{}From the properties of $Z_{k}$ we have that $\Gamma(k) < 1$ for any value
of $k$ and of the parameters $m_{1}$ and $m_{2}$. By using the customary
continuous limit relation $\sum_{k}\;\rightarrow \;
\frac{V}{(2\pi)^{3}}\int d^{3}k$, in the infinite volume limit we obtain
$$\lim_{V \rightarrow \infty}\; _{1,2}\langle0|0\rangle_{e,\mu} =
\lim_{V \rightarrow \infty}\; e^{\frac{V}{(2\pi)^{3}}\int d^{3}k
\;ln\; \Gamma(k)}= 0 \eqno(2.34)$$
Of course, this orthogonality disappears when
$\theta =0$ and/or when $m_{1} = m_{2}$ (because in this case $Z_{k}=0$
and no mixing occurs in Pontecorvo theory).

Eq.(2.34) expresses the unitary inequivalence in the infinite volume limit
of the flavor and the mass representations and shows
the absolutely non-trivial nature of the mixing
transformations (2.1). In other words, the mixing transformations induce a
physically non-trivial structure in the flavor vacuum which indeed turns out
to be an $SU(2)$ generalized coherent state. In Section 4
we will see how such a vacuum structure may lead to
phenomenological consequences in the neutrino oscillations, which possibly
may be experimentally tested. From eq.(2.34) we also see that eq.(2.22) is a
purely formal expression which only holds at finite volume.

We thus realize the
limit of validity of the approximation usually adopted when the mass vacuum
state (representation for definite mass operators) is identified
with the vacuum for the flavor operators.
We point out that even at finite volume the vacua identification is actually
an approximation since the flavor vacuum is an $SU(2)$ generalized coherent
state. In such an approximation, the coherent state structure and
many physical features are missed.

\bigskip
\bigskip

{\bf 3  The number operator, the dynamical map and the mass sectors}

\medskip

We now calculate the number of the particles condensed in the state
$|0\rangle_{e,\mu}$.

Let us consider, for example, the $\alpha_{k,1}$ particles.
As usual we define the number operator
as $N^{k}_{\alpha_{1}}\equiv \sum_{r}\alpha^{r \dag}_{k,1}\alpha^{r}_{k,1}$
and use the fact that $N^{k}_{\alpha_{1}}$ commutes with $S_{3}^{p}$ and with
$S_{\pm}^{p}$ for $p \neq k$. Then we can write
$$_{e,\mu}\langle0|N^{k}_{\alpha_{1}}|0\rangle_{e,\mu}=\;
^{\;\;k}_{e,\mu}\langle0|N^{k}_{\alpha_{1}}|0\rangle_{e,\mu}^{k} \eqno(3.1) $$
$|0\rangle_{e,\mu}^{k}$ has been introduced in eq.(2.28).
{}From eq.(3.1) and the relations given in Appendix C we obtain:
$$\;_{e,\mu}\langle0|N^{k}_{\alpha_{1}}|0\rangle_{e,\mu}=Z_{k}\sin^{2}\theta
\;.
\eqno(3.2)$$
The same result is obtained for the number
operators $N^{k}_{\alpha_{2}}$, $N^{k}_{\beta_{1}}$, $N^{k}_{\beta_{2}}$:
$$ \;_{e,\mu}\langle0|N^{k}_{\sigma_{i}}|0\rangle_{e,\mu}
=Z_{k}\sin^{2}\theta \;\;\;,\;\;\; \sigma=\alpha,\beta\;,\; i=1,2\;.
\eqno(3.3)$$

Eq.(3.2) gives the condensation density of the flavor vacuum state
as a function of the mixing angle $\theta$, of the
masses $m_{1}$ and $m_{2}$, and of the momentum modulus $k$. This last
feature is particularly interesting since, as we will see, the
vacuum acts as a "momentum (or spectrum) analyzer" when time-evolution and
flavor oscillations are considered.

We remark that the eq.(3.2) (and (3.3)) clearly shows that the flavor vacuum
$|0\rangle_{e,\mu}$ is not annihilated by the operators $\alpha^{r}_{k,i}$,
$\beta^{r}_{k,i}$, $i=1,2$. This is in contrast with the usual treatment where
the flavor vacuum $|0\rangle_{e,\mu}$
is identified with the mass vacuum $|0\rangle_{1,2}$.

Notice that, due to the behaviour of $Z_{k}$ for high $k$, expectation
values of $N^{k}_{\sigma_{i}}$ are zero for high $k$ (the same is true
for any operator $O_{i}^{k}$, $i=1,2$, for which is
$\;_{1,2}\langle0|O^{k}_{i}|0\rangle_{1,2}=0  $) .

In order to explicitly exhibit the dynamical map, eqs.(2.23),
it is convenient to redefine the operatorial parts of the fields
$\nu_{e}(x)$ and $\nu_{\mu}(x)$ as (cf. eqs.(2.23))
$u_{k,1}^{r,\alpha}\alpha_{k,e}^{r} \equiv u_{k,e}^{r,\alpha}
{\tilde\alpha}_{k,e}^{r}$, etc., so that we can write:
$$\alpha^{r}_{k,e} \equiv G^{-1}(\theta) \;
\alpha^{r}_{k,1}\;G(\theta)\;,\eqno(3.4a)$$
$$\alpha^{r}_{k,\mu} \equiv G^{-1}(\theta)\;
\alpha^{r}_{k,2}\;G(\theta)\;,\eqno(3.4b)$$
$$ \;\beta^{r}_{k,e} \equiv
G^{-1}(\theta) \;\beta^{r}_{k,1}\;G(\theta)\;,\eqno(3.4c)$$
$$\beta^{r}_{k,\mu} \equiv
G^{-1}(\theta) \;\beta^{r}_{k,2}\;G(\theta)\;.\eqno(3.4d)$$

We observe that $\alpha^{r}_{k,l} $ and $\beta^{r}_{k,l}$,
$l=e,\mu$, depend on time through the time dependence of
$G(\theta)$. We obtain:

$$\alpha^{r}_{k,e}=\cos\theta\;\alpha^{r}_{k,1}\;+\;\sin\theta\;\sum_{s}\left(
(u^{r\dag}_{k,1} u^{s}_{k,2})\; \alpha^{s}_{k,2}\;+\;
(u^{r\dag}_{k,1} v^{s}_{-k,2})\; \beta^{s\dag}_{-k,2}\right)\eqno(3.5a)$$
%
$$\alpha^{r}_{k,\mu}=\cos\theta\;\alpha^{r}_{k,2}\;-\;\sin\theta\;\sum_{s}\left(
(u^{r\dag}_{k,2} u^{s}_{k,1})\; \alpha^{s}_{k,1}\;+\;
(u^{r\dag}_{k,2} v^{s}_{-k,1})\; \beta^{s\dag}_{-k,1}\right)\eqno(3.5b)$$
$$\beta^{r}_{-k,e}=\cos\theta\;\beta^{r}_{-k,1}\;+\;\sin\theta\;\sum_{s}\left(
(v^{s\dag}_{-k,2} v^{r}_{-k,1})\; \beta^{s}_{-k,2}\;+\;
(u^{s\dag}_{k,2} v^{r}_{-k,1})\; \alpha^{s\dag}_{k,2}\right)\eqno(3.5c)$$
%
$$\beta^{r}_{-k,\mu}=\cos\theta\;\beta^{r}_{-k,2}\;-\;\sin\theta\;\sum_{s}\left(
(v^{s\dag}_{-k,1} v^{r}_{-k,2})\; \beta^{s}_{-k,1}\;+\;
(u^{s\dag}_{k,1} v^{r}_{-k,2})\; \alpha^{s\dag}_{k,1}\right)\eqno(3.5d)$$

Without loss of generality, we can choose  the reference frame
such that $k=(0,0,|k|)$.
This implies that only the products of wave functions with $r=s$
will survive (see Appendix B). Eqs.(3.5) then assume the simpler
form:
$$\alpha^{r}_{k,e}=\cos\theta\;\alpha^{r}_{k,1}\;+\;\sin\theta\;\left(
U_{k}^{*}\; \alpha^{r}_{k,2}\;+\;\epsilon^{r}\;
V_{k}\; \beta^{r\dag}_{-k,2}\right) \eqno(3.6a)$$
$$\alpha^{r}_{k,\mu}=\cos\theta\;\alpha^{r}_{k,2}\;-\;\sin\theta\;\left(
U_{k}\; \alpha^{r}_{k,1}\;-\;\epsilon^{r}\;
V_{k}\; \beta^{r\dag}_{-k,1}\right)  \eqno(3.6b)$$
$$\beta^{r}_{-k,e}=\cos\theta\;\beta^{r}_{-k,1}\;+\;\sin\theta\;\left(
U_{k}^{*}\; \beta^{r}_{-k,2}\;-\;\epsilon^{r}\;
V_{k}\; \alpha^{r\dag}_{k,2}\right) \eqno(3.6c)$$
$$\beta^{r}_{-k,\mu}=\cos\theta\;\beta^{r}_{-k,2}\;-\;\sin\theta\;\left(
U_{k}\; \beta^{r}_{-k,1}\;+\;\epsilon^{r}\;
V_{k}\; \alpha^{r\dag}_{k,1}\right) \eqno(3.6d)$$
with $\epsilon^{r}=(-1)^{r}$ and
$$U_{k}\equiv(u^{r\dag}_{k,2}u^{r}_{k,1})=
(v^{r\dag}_{-k,1}v^{r}_{-k,2}) \eqno(3.7a)$$
$$V_{k}\equiv \epsilon^{r}\;(u^{r\dag}_{k,1}v^{r}_{-k,2})=
-\epsilon^{r}\;(u^{r\dag}_{k,2}v^{r}_{-k,1}) \eqno(3.7b)$$
where the time dependence of $U_{k}$ and $V_{k}$ has been omitted. We have:
$$V_{k}=|V_{k}|\;e^{i(\omega_{k,2}+\omega_{k,1})t}\;\;\;\;,\;\;\;\;
U_{k}=|U_{k}|\;e^{i(\omega_{k,2}-\omega_{k,1})t} \eqno(3.8)$$
$$|U_{k}|=\left(\frac{\omega_{k,1}+m_{1}}{2\omega_{k,1}}\right)^{\frac{1}{2}}
\left(\frac{\omega_{k,2}+m_{2}}{2\omega_{k,2}}\right)^{\frac{1}{2}}
\left(1+\frac{k^{2}}{(\omega_{k,1}+m_{1})(\omega_{k,2}+m_{2})}\right)
\eqno(3.9a)$$
$$|V_{k}|=\left(\frac{\omega_{k,1}+m_{1}}{2\omega_{k,1}}\right)^{\frac{1}{2}}
\left(\frac{\omega_{k,2}+m_{2}}{2\omega_{k,2}}\right)^{\frac{1}{2}}
\left(\frac{k}{(\omega_{k,2}+m_{2})}-\frac{k}{(\omega_{k,1}+m_{1})}\right)
\eqno(3.9b)$$
$$|U_{k}|^{2}+|V_{k}|^{2}=1 \;\;\;\;\;,
\;\;\;\;\;|V_{k}|^{2}=\frac{1}{2}\;Z_{k} \eqno(3.10)$$
For notational simplicity in the following we put
$\omega_{i}\equiv\omega_{k,i}$.

It is also interesting to exhibit the explicit expression of
$|0\rangle_{e,\mu}^{k}$ in the reference frame for which $k=(0,0,|k|)$ (see
Appendix D):
$$|0\rangle_{e,\mu}^{k}= \prod_{r} \left[(1-\sin^{2}\theta\;|V_{k}|^{2})
-\epsilon^{r}\sin\theta\;\cos\theta\; V_{k} (A^{r}+B^{r})+
\right. $$
$$\left.
+\;\epsilon^{r}\sin^{2}\theta \;V_{k} (U_{k}^{*} C^{r}- U_{k} D^{r})
+\sin^{2}\theta \; V_{k}^{2} A^{r}B^{r} \right]|0\rangle_{1,2}
\eqno(3.11)$$
with
$$A_{k}^{r}\equiv\alpha^{r\dag}_{k,1}\beta^{r\dag}_{-k,2}\;\;\;,\;\;\;
  B_{k}^{r}\equiv\alpha^{r\dag}_{k,2}\beta^{r\dag}_{-k,1}\;\;\;,\;\;\;
  C_{k}^{r}\equiv\alpha^{r\dag}_{k,1}\beta^{r\dag}_{-k,1}\;\;\;,\;\;\;
  D_{k}^{r}\equiv\alpha^{r\dag}_{k,2}\beta^{r\dag}_{-k,2}\eqno(3.12)$$

We observe that eqs.(3.6) can be obtained by a rotation and by a subsequent
Bogoliubov transformation. To see this it is convenient to put (cf. eq.(3.10)):
$$|U_{k}|\equiv \cos\Theta_{k} \;\;\;\;\;\;,
\;\;\;\;\;\; |V_{k}|\equiv \sin\Theta_{k} \;\;\;\;\;\;,
\;\;\;\;\;\; 0 \leq \Theta_{k} <\frac{\pi}{4} \eqno(3.13)$$
and
$$e^{i(\omega_{1}-\omega_{2})t} \equiv e^{i\psi} \;\;\;\;\;\;,
\;\;\;\;\;\; e^{2i\omega_{1}t}\equiv e^{i\phi_{1}} \;\;\;\;\;\;,
\;\;\;\;\;\; e^{2i\omega_{2}t}\equiv e^{i\phi_{2}}  \eqno(3.14)$$
so that eqs.(3.6) are rewritten as
$$ \alpha_{k,e}^{r}=B_{2}^{-1}R^{-1}\alpha_{k,1}^{r}RB_{2} \eqno(3.15a)$$
$$ \beta_{-k,e}^{r}=B_{2}^{-1}R^{-1}\beta_{-k,1}^{r}RB_{2} \eqno(3.15b)$$
$$ \alpha_{k,\mu}^{r}=B_{1}^{-1}R^{-1}\alpha_{k,2}^{r}RB_{1} \eqno(3.15c)$$
$$ \beta_{-k,\mu}^{r}=B_{1}^{-1}R^{-1}\beta_{-k,2}^{r}RB_{1} \eqno(3.15d)$$
where
$$R=exp\left\{ \theta\; \sum_{k,r}\left[
\left( \alpha_{k,1}^{r\dag}\alpha_{k,2}^{r}+
\beta_{-k,1}^{r\dag}\beta_{-k,2}^{r} \right)e^{i\psi}
-\left( \alpha_{k,2}^{r\dag}\alpha_{k,1}^{r}+
\beta_{-k,2}^{r\dag}\beta_{-k,1}^{r} \right)e^{-i\psi}
\right]    \right\}\eqno(3.16)$$
$$B_{1}=exp\left\{ -\;\sum_{k,r}\;\Theta_{k}\;\epsilon^{r}\; \left[
\alpha_{k,1}^{r}\beta_{-k,1}^{r}\;e^{-i\phi_{1}}
-\beta_{-k,1}^{r\dag}\alpha_{k,1}^{r\dag}\;e^{i\phi_{1}}
\right]    \right\}\eqno(3.17)$$
$$B_{2}=exp\left\{ \sum_{k,r}\;\Theta_{k}\;\epsilon^{r}\; \left[
\alpha_{k,2}^{r}\beta_{-k,2}^{r}\;e^{-i\phi_{2}}
-\beta_{-k,2}^{r\dag}\alpha_{k,2}^{r\dag}\;e^{i\phi_{2}}
\right]    \right\}\eqno(3.18)$$

By use of these relations
and noting that $R|0\rangle_{1,2}=|0\rangle_{1,2}$, we can
"separate" the sectors $\left\{ |0(\Theta)\rangle_{1}\right\}$ and
$\left\{ |0(\Theta)\rangle_{2}\right\}$ out of the full representation space
$\left\{ |0\rangle_{e,\mu}\right\}:$ $\left\{ |0(\Theta)\rangle_{1}\right\}
\otimes \left\{ |0(\Theta)\rangle_{2}\right\}
\subset \left\{ |0\rangle_{e,\mu}\right\}$.

The states $|0(\Theta)\rangle_{1}$ and
$|0(\Theta)\rangle_{2}$ are respectively obtained as:
$$|0(\Theta)\rangle_{1}\equiv B^{-1}_{1}(\Theta) |0\rangle_{1}=
\prod_{k,r}\left(\cos\Theta_{k}+
\epsilon^{r}\;e^{i\phi_{1}}\;\sin\Theta_{k}\;\beta_{-k,1}^{r\dag}
\alpha_{k,1}^{r\dag}
\right) |0\rangle_{1}\eqno(3.19)$$
$$|0(\Theta)\rangle_{2}\equiv B^{-1}_{2}(\Theta) |0\rangle_{2}=
\prod_{k,r}\left(\cos\Theta_{k}-
\epsilon^{r}\;e^{i\phi_{2}}\;\sin\Theta_{k}\;\beta_{-k,2}^{r\dag}
\alpha_{k,2}^{r\dag}
\right) |0\rangle_{2}\eqno(3.20)$$

If one wants to work
with the "mass" sectors $\left\{ |0(\Theta)\rangle_{1}\right\}$ and
$\left\{ |0(\Theta)\rangle_{2}\right\}$, the tensor product formalism must be
used, e.g.
$$\left(O_{1} + O_{2}\right) \left(|0(\Theta)\rangle_{1}\otimes
|0(\Theta)\rangle_{2}\right)\equiv
\left(O_{1} \otimes I + I \otimes O_{2}\right) \left(|0(\Theta)\rangle_{1}
 \otimes
|0(\Theta)\rangle_{2}\right)=$$
$$=O_{1}|0(\Theta)\rangle_{1} \otimes |0(\Theta)\rangle_{2}+
|0(\Theta)\rangle_{1} \otimes O_{2}|0(\Theta)\rangle_{2} \eqno(3.21)$$
with $O_{i}$, $i=1,2$, any product of $\nu_{i}$ neutrino
field operators. For example, we have
$|0\rangle_{1,2}\equiv|0\rangle_{1}\otimes |0\rangle_{2}$,
 $\alpha_{k,1}^{r}\equiv\alpha_{k,1}^{r} \otimes I$,
 $\alpha_{k,2}^{r}\equiv I \otimes\alpha_{k,2}^{r}$, so that
$\alpha_{k,1}^{r}\alpha_{k,2}^{r\dag}=
\left(\alpha_{k,1}^{r}\otimes I \right)
\left(I \otimes\alpha_{k,2}^{r\dag}\right)=
\alpha_{k,1}^{r} \otimes\alpha_{k,2}^{r\dag}$ and
$$R^{-1}\alpha_{k,1}^{r}R =\cos \theta
\left(\alpha_{k,1}^{r}\otimes I\right)+
e^{i\psi}\;\sin \theta\left(I \otimes\alpha_{k,2}^{r}\right) \eqno(3.22)$$
$$B_{2}^{-1}R^{-1}\alpha_{k,1}^{r}R B_{2}=\cos \theta
\left(\alpha_{k,1}^{r}\otimes I\right)+
e^{i\psi}\;\sin \theta\left(I \otimes\alpha_{k,2}^{r}(\Theta)\right)
\eqno(3.23)$$
with
$$\alpha_{k,2}^{r}(\Theta)=\cos\Theta_{k}\; \alpha_{k,2}^{r} +
\epsilon^{r}\;e^{i\phi_{2}}
\;\sin\Theta_{k}\; \beta^{r\dag}_{-k,2} \eqno(3.24) $$
and
$$B_{2}^{-1}|0\rangle_{1,2}
=\left(|0\rangle_{1} \otimes |0(\Theta)\rangle_{2}\right)\;. \eqno(3.25)$$

We note that $|0(\Theta)\rangle_{i}$, $i=1,2$, are the vacuum states for
$\alpha_{k,i}^{r}(\Theta)=B(\Theta)_{i}^{-1} \alpha_{k,i}^{r}B(\Theta)_{i} $
and $\beta_{k,i}^{r}(\Theta)=B(\Theta)_{i}^{-1} \beta_{k,i}^{r}B(\Theta)_{i} $
operators.

We also note
that $\alpha_{k,e}^{r}
\left(|0\rangle_{1} \otimes |0(\Theta)\rangle_{2}\right)
=\beta_{k,e}^{r} \left(|0\rangle_{1} \otimes |0(\Theta)\rangle_{2}\right)
=0$;  $\alpha_{k,\mu}^{r}
\left(|0(\Theta)\rangle_{1} \otimes |0\rangle_{2}\right)
=\beta_{k,\mu}^{r}
\left(|0(\Theta)\rangle_{1} \otimes |0\rangle_{2}\right)
=0$, but
$\alpha_{k,e}^{r}
\left(|0(\Theta)\rangle_{1} \otimes |0\rangle_{2}\right)\neq0$,
  $\alpha_{k,\mu}^{r}
\left(|0\rangle_{1} \otimes |0(\Theta)\rangle_{2}\right)\neq0$, etc..
Moreover, $\;_{2}\langle0(\Theta)|N_{\sigma_{2}}^{k,r}
|0(\Theta)\rangle_{2}= \sin^{2}\Theta_{k}$,
 $\;_{1}\langle0(\Theta)|N_{\sigma_{1}}^{k,r}
|0(\Theta)\rangle_{1}= \sin^{2}\Theta_{k}$,  and
$\;_{2}\langle0(\Theta)|N_{\sigma_{1}}^{k,r}
|0(\Theta)\rangle_{2}=\;_{1}\langle0(\Theta)|N_{\sigma_{2}}^{k,r}
|0(\Theta)\rangle_{1}= 0$,  $\sigma=\alpha,\beta$, which show the condensate
structure of the sectors $\left\{|0(\Theta)\rangle_{i}\right\}\;,\;\;i=1,2$.

Finally, we observe that $|0(\Theta)\rangle_{i}$
can be written as
$$|0(\Theta)\rangle_{i}= exp\left(-\frac{S_{\alpha_{i}}}{2}\right)
|I_{i}\rangle=exp\left(-\frac{S_{\beta_{i}}}{2}\right)
|I_{i}\rangle \eqno(3.26)$$
with $|I_{i}\rangle\equiv
exp\left(\sum_{k,r}(-1)^{i+1}\epsilon^{r}e^{i\phi_{i}}
\beta_{-k,i}^{r\dag}\alpha_{k,i}^{r\dag}\right)|0\rangle_{1,2}$, and
$$S_{\alpha_{i}}= -\sum_{k,r}\left(\alpha_{k,i}^{r\dag}\alpha_{k,i}^{r}
\log\sin^{2}\Theta_{k} + \alpha_{k,i}^{r}\alpha_{k,i}^{r\dag}
\log\cos^{2}\Theta_{k}\right)\;\;\;,\;\;\; i=1,2\;. \eqno(3.27)$$
A similar expression holds for $S_{\beta_{i}}$.
It is known that $S_{\alpha_{i}}$
(or $S_{\beta_{i}}$) can be interpreted as the entropy function associated to
the vacuum condensate [7].


\bigskip
\bigskip

{\bf 4  Neutrino oscillations}

\medskip

We are now ready to study the flavor oscillations. In order to compare our
result with the conventional one [1,2,5], we first reproduce the usual
oscillation formula.

In the original Pontecorvo and collaborators treatment [5],
the vacuum state for definite flavor neutrinos is identified with the vacuum
state for definite mass neutrinos: $|0\rangle_{e,\mu}=|0\rangle_{1,2}
\equiv|0\rangle$.
As we have shown in the previous Section, such an identification is
not possible in QFT; however, it is allowed at finite volume where no problem
of unitary inequivalence
arises in the choice of the Hilbert space. As already observed, even at finite
volume, the vacua identification is only an approximation. For shortness we
refer to such an identification simply as to the finite volume approximation,
meaning by that the approximation which is allowed at finite volume.

The number operators relative to electronic and muonic
neutrinos are
$$N_{\alpha_{e}}^{k,r}=\alpha_{k,e}^{r\dag}\alpha_{k,e}^{r}= $$
$$=\cos^{2}\theta \;\alpha_{k,1}^{r\dag}\alpha_{k,1}^{r} +
\sin^{2}\theta \;\alpha_{k,2}^{r\dag}\alpha_{k,2}^{r}+
\sin\theta \cos\theta\;(\alpha_{k,1}^{r\dag}\alpha_{k,2}^{r}
+ \alpha_{k,2}^{r\dag}\alpha_{k,1}^{r})\;, \eqno(4.1)$$
$$N_{\alpha_{\mu}}^{k,r}=\alpha_{k,\mu}^{r\dag}\alpha_{k,\mu}^{r}=$$
$$=\cos^{2}\theta \alpha_{k,2}^{r\dag}\alpha_{k,2}^{r} +
\sin^{2}\theta \alpha_{k,1}^{r\dag}\alpha_{k,1}^{r}-
\sin\theta \cos\theta(\alpha_{k,1}^{r\dag}\alpha_{k,2}^{r}
+ \alpha_{k,2}^{r\dag}\alpha_{k,1}^{r})\;, \eqno(4.2)$$
and, obviously,
$$\langle0|\;N_{\alpha_{e}}^{k,r}\;
|0\rangle\;=\;\langle0|\;N_{\alpha_{\mu}}^{k,r}\;
|0\rangle\;=0\;. \eqno(4.3)$$

The one electronic neutrino state is of the form
$$|\alpha_{k,e}^{r}\rangle\;=\;
\cos\theta\; |\alpha_{k,1}^{r}\rangle + \sin\theta\; |\alpha_{k,2}^{r}\rangle.
\eqno(4.4)$$
where
$|\alpha^{r}_{k,e}\rangle \equiv\alpha^{r\dag}_{k,e}|0\rangle
\;,\;\;\;|\alpha^{r}_{k,1}\rangle \equiv\alpha^{r\dag}_{k,1}|0\rangle
\;,\;\;\;|\alpha^{r}_{k,2}\rangle \equiv\alpha^{r\dag}_{k,2}|0\rangle $.
The time evoluted of this state is controlled by the time evolution of
$|\alpha^{r}_{k,1}\rangle $ and $|\alpha^{r}_{k,2}\rangle$
$$|\alpha_{k,e}^{r}(t)\rangle\;=\;e^{-iH_{1,2}t}
|\alpha_{k,e}^{r}\rangle\;=\;
\cos\theta\;e^{-i\omega_{1} t} |\alpha_{k,1}^{r}\rangle +
\sin\theta\;e^{-i\omega_{2} t}|\alpha_{k,2}^{r}\rangle\;. \eqno(4.5)$$

We have
$$\langle\alpha_{k,e}^{r}|\;N_{\alpha_{e}}^{k,r}\;
|\alpha_{k,e}^{r}\rangle\;=\;1 \eqno(4.6)$$
and
$$\langle\alpha_{k,e}^{r}(t)|\;N_{\alpha_{e}}^{k,r}\;
|\alpha_{k,e}^{r}(t)\rangle\;=\;
\cos^{4}\theta + \sin^{4}\theta +
\sin^{2}\theta\;\cos^{2}\theta (e^{-i(\omega_{2}-\omega_{1})
t}+e^{+i(\omega_{2}-\omega_{1}) t}) =$$
$$=1-\sin^{2}2\theta\;\sin^{2}\left(\frac{\Delta\omega}{2}t
\right)\;.  \eqno(4.7)$$
The number of $\alpha_{e}$ particles therefore
oscillates in time with a frequency
given by the difference $\Delta\omega$ in the energies of the physical
components $\alpha_{1}$ and $\alpha_{2}$. This oscillation is a flavor
oscillation since we have at the same time:
$$\langle\alpha_{k,e}^{r}(t)|\;N_{\alpha_{\mu}}^{k,r}\;
|\alpha_{k,e}^{r}(t)\rangle\;=\;
\sin^{2}2\theta\;\sin^{2}\left(\frac{\Delta\omega}{2}t
\right) \eqno(4.8)$$
so that
$$\langle\alpha_{k,e}^{r}(t)|\;N_{\alpha_{e}}^{k,r}
\;|\alpha_{k,e}^{r}(t)\rangle\;+
\langle\alpha_{k,e}^{r}(t)|\; N_{\alpha_{\mu}}^{k,r}
\;|\alpha_{k,e}^{r}(t)\rangle\;=\;1\;. \eqno(4.9)$$
Notice that the traditional derivation of eq.(4.7) is the same as the one
presented above since
$\;\langle\alpha_{k,e}^{r}(t)|\;N_{\alpha_{e}}^{k,r}\;
|\alpha_{k,e}^{r}(t)\rangle=|\langle\alpha_{k,e}^{r}|
\alpha_{k,e}^{r}(t)\rangle|^{2}$, as can be seen from eqs.(4.4) and (4.5).
Eqs.(4.7)-(4.9) are the well known results.

Let us now go to the QFT framework.

We have seen that $|0\rangle_{1,2}$ and
$|0\rangle_{e,\mu}$ are orthogonal in the infinite volume limit. We choose to
work in the physical (mass) representation {$|0\rangle_{1,2}$} (same
conclusions
are of course reached by working, with due changes, in the flavor
representation
{$|0\rangle_{e,\mu}$} ) in order to follow the time evolution of the physical
components $|\alpha^{r}_{k,1}\rangle$ and $|\alpha^{r}_{k,2}\rangle$ .

The number operators are now (cf. eqs.(3.6))
$$N_{\alpha_{e}}^{k,r}=\alpha_{k,e}^{r\dag}\alpha_{k,e}^{r}=$$
$$=\cos^{2}\theta \alpha_{k,1}^{r\dag}\alpha_{k,1}^{r} +
\sin^{2}\theta\;|U_{k}|^{2}\; \alpha_{k,2}^{r\dag}\alpha_{k,2}^{r}
+\sin^{2}\theta\;|V_{k}|^{2}\; \beta_{k,2}^{r}\beta_{k,2}^{r\dag}+$$
$$+\sin\theta \cos\theta \left(U_{k}^{*}\;\alpha_{k,1}^{r\dag}\alpha_{k,2}^{r}
+ \;U_{k}\;\alpha_{k,2}^{r\dag}\alpha_{k,1}^{r}
+\epsilon^{r}\;V_{k}^{*}\;\beta_{k,2}^{r}\alpha_{k,1}^{r}
+\epsilon^{r}\;V_{k}\;\alpha_{k,1}^{r\dag}\beta_{k,2}^{r\dag}\right)+$$
$$+\epsilon^{r}\sin^{2}\theta\;
\left(V_{k}\;U_{k}\; \alpha_{k,2}^{r\dag}\beta_{k,2}^{r\dag}
+V_{k}^{*}\;U_{k}^{*}\; \beta_{k,2}^{r}\alpha_{k,2}^{r}\right) \;,
\eqno(4.10)$$
$$N_{\alpha_{\mu}}^{k,r}=\alpha_{k,\mu}^{r\dag}\alpha_{k,\mu}^{r}=$$
$$=\cos^{2}\theta \alpha_{k,2}^{r\dag}\alpha_{k,2}^{r} +
\sin^{2}\theta\;|U_{k}|^{2}\; \alpha_{k,1}^{r\dag}\alpha_{k,1}^{r}
+\sin^{2}\theta\;|V_{k}|^{2}\; \beta_{k,1}^{r}\beta_{k,1}^{r\dag}+$$
$$-\sin\theta \cos\theta \left(U_{k}^{*}\;\alpha_{k,1}^{r\dag}\alpha_{k,2}^{r}
+ \;U_{k}\;\alpha_{k,2}^{r\dag}\alpha_{k,1}^{r}
-\epsilon^{r}\;V_{k}^{*}\;\beta_{k,1}^{r}\alpha_{k,2}^{r}
-\epsilon^{r} \;V_{k}\;\alpha_{k,2}^{r\dag}\beta_{k,1}^{r\dag}\right)+$$
$$-\epsilon^{r}\;\sin^{2}\theta\;
\left(V_{k}\;U_{k}^{*}\; \alpha_{k,1}^{r\dag}\beta_{k,1}^{r\dag}
+V_{k}^{*}\;U_{k}\; \beta_{k,1}^{r}\alpha_{k,1}^{r}\right)\;. \eqno(4.11)$$

The one electronic neutrino state is given by
$$|\alpha^{r}_{k,e}\rangle \equiv\alpha^{r\dag}_{k,e}|0\rangle_{1,2}
=\cos\theta\;|\alpha^{r}_{k,1}\rangle+\;
\sin\theta\;U_{k}\; |\alpha^{r}_{k,2}\rangle \eqno(4.12)$$
with $|\alpha^{r}_{k,i}\rangle
\equiv\alpha^{r\dag}_{k,i}|0\rangle_{1,2}$, $i=1,2$.
The action of the flavor operator $\alpha^{r\dag}_{k,e}$
is defined on $|0\rangle_{1,2}$ through
the mapping (3.6) (see also eqs.(3.4)). Note that, as it must be,
$$\langle\alpha_{k,e}^{r}|\alpha_{k,e}^{r}\rangle\;=
\;1 -\;\sin^{2}\theta\;|V_{k}|^{2}=
1\;-\;_{1,2}\langle0|\;\alpha_{k,e}^{r\dag}\alpha_{k,e}^{r}\;
|0\rangle_{1,2} \;, \eqno(4.13)$$
since $\alpha_{k,e}^{r}\;|0\rangle_{1,2} \neq 0$.

The time evolution of the state $|\alpha_{k,e}^{r}\rangle$ is given by
$$|\alpha^{r}_{k,e}(t)\rangle
=e^{-i H_{1,2} t} |\alpha_{k,e}^{r}\rangle =\cos\theta\;e^{-i\omega_{1} t}
|\alpha_{k,1}^{r}\rangle +
\sin\theta\;U_{k}\;e^{-i\omega_{2} t}|\alpha_{k,2}^{r}\rangle \;. \eqno(4.14)$$
We have:
$$\langle\alpha_{k,e}^{r}|\;N_{\alpha_{e}}^{k,r}\;
|\alpha_{k,e}^{r}\rangle\;=\;1 -\;\sin^{2}\theta\;|V_{k}|^{2}=
1\;-\;_{1,2}\langle0|\;N_{\alpha_{e}}^{k,r}\;
|0\rangle_{1,2} \eqno(4.15)$$
so that,
$$\langle\alpha_{k,e}^{r}|\;N_{\alpha_{e}}^{k,r}\;|\alpha_{k,e}^{r}\rangle\;+
\;_{1,2}\langle0|\;N_{\alpha_{e}}^{k,r}\;|0\rangle_{1,2}\;=\;1\;. \eqno(4.16)$$
We note that the expectation value of  $N_{\alpha_{e}}^{k,r}$ in the vacuum
$|0\rangle_{1,2}$ provides an essential contribution to the normalization
equation (4.16) (see also eq.(4.13)).

It is also interesting to observe that the term
$\;_{1,2}\langle0|\;N_{\alpha_{e}}^{k,r}\;|0\rangle_{1,2}\;$ plays the r\^ole
of zero point contribution when considering the energy contribution
of $\alpha_{e}^{k,r}$ particles.

Note that we also have:
$$\;_{1,2}\langle0|\;N_{\sigma_{l}}^{k,r}\;|0\rangle_{1,2}\;=
\;\sin^{2}\theta\;|V_{k}|^{2} \;,\;\;\;\; \sigma=\alpha, \beta \;,\;\;\;\;
l=e,\mu ,\eqno(4.17)$$
and
$$\langle\alpha_{k,e}^{r}|\;N_{\alpha_{\mu}}^{k,r}\;|\alpha_{k,e}^{r}\rangle\;
= \;\sin^{2}\theta\;|V_{k}|^{2}\;
\left(1\;-\;\sin^{2}\theta\;|V_{k}|^{2}\right)\;. \eqno(4.18)$$
Eqs.(4.17) show the condensate structure of $|0\rangle_{1,2}$ in terms of
definite flavor fields and of course are analogous to eqs.(3.3) which
show the condensate structure of $|0\rangle_{e,\mu}$ in terms of
definite mass fields.

We also have
$$\langle\alpha_{k,e}^{r}(t)|\;N_{\alpha_{e}}^{k,r}\;
|\alpha_{k,e}^{r}(t)\rangle\;=$$
$$=\cos^{4}\theta + |U_{k}|^{2}\;\sin^{4}\theta +
|V_{k}|^{2}\;\sin^{2}\theta\;\cos^{2}\theta
+ |U_{k}|^{2}\;\sin^{2}\theta\;\cos^{2}\theta (e^{-i(\omega_{2}-\omega_{1})
t}+e^{+i(\omega_{2}-\omega_{1}) t})= $$
$$=\left(\;1 -\;\sin^{2}\theta\;|V_{k}|^{2}\right)
-\;|U_{k}|^{2}\;\sin^{2}2\theta
\;\sin^{2}\left(\frac{\Delta\omega}{2}t\right)\;. \eqno(4.19)$$
This result reproduces the one obtained in the finite volume approximation
(cf. eq.(4.7)) when $|U_{k}|\rightarrow 1$ (and $|V_{k}|\rightarrow 0$).
The fraction of $\alpha_{\mu}^{k,r}$ particles in the same
state is
$$\langle\alpha_{k,e}^{r}(t)|\;N_{\alpha_{\mu}}^{k,r}\;
|\alpha_{k,e}^{r}(t)\rangle\;=$$
$$=\;|U_{k}|^{2}\;\sin^{2}2\theta
\;\sin^{2}\left(\frac{\Delta\omega}{2}t\right)+
\;\sin^{2}\theta\;|V_{k}|^{2}\;
\left(1\;-\;\sin^{2}\theta\;|V_{k}|^{2}\right)\;,\eqno(4.20)$$
where we recognize the contribution from the $\alpha_{k,\mu}^{r}$ condensate
in the state $|\alpha_{k,e}^{r}\rangle$ (cf. eq.(4.18)). Eq.(4.19) is to
be compared with the approximated one (4.7).
Note that
$$\langle\alpha_{k,e}^{r}(t)|\;N_{\alpha_{e}}^{k,r}\;
|\alpha_{k,e}^{r}(t)\rangle +
\langle\alpha_{k,e}^{r}(t)|\;N_{\alpha_{\mu}}^{k,r}
\;|\alpha_{k,e}^{r}(t)\rangle=
\langle\alpha_{k,e}^{r}|\;N_{\alpha_{e}}^{k,r}\;
|\alpha_{k,e}^{r}\rangle +
\langle\alpha_{k,e}^{r}|\;N_{\alpha_{\mu}}^{k,r}
\;|\alpha_{k,e}^{r}\rangle\;. \eqno(4.21)$$
The normalization relation is written as
$$\langle\alpha_{k,e}^{r}(t)|\;N_{\alpha_{e}}^{k,r}\;
|\alpha_{k,e}^{r}(t)\rangle +
\;_{1,2}\langle0|\;N_{\alpha_{e}}^{k,r}\;|0\rangle_{1,2}\;
+\langle\alpha_{k,e}^{r}(t)|\;N_{\alpha_{\mu}}^{k,r}
\;|\alpha_{k,e}^{r}(t)\rangle\;
-\;\langle\alpha_{k,e}^{r}|\;N_{\alpha_{\mu}}^{k,r}\;
|\alpha_{k,e}^{r}\rangle=\;1\;. \eqno(4.22)$$

We may also use ${\cal N}_{k}\equiv \langle\alpha_{k,e}^{r}
|\alpha_{k,e}^{r}\rangle$ as normalization factor and write eqs.(4.15), (4.18),
(4.19), (4.20) and the normalization relation as

$$\frac{1}{{\cal N}_{k}}\langle\alpha_{k,e}^{r}|\;N_{\alpha_{e}}^{k,r}\;
|\alpha_{k,e}^{r}\rangle\;=\;1 \eqno(4.23)$$
$$\frac{1}{{\cal N}_{k}}\langle\alpha_{k,e}^{r}|
\;N_{\alpha_{\mu}}^{k,r}\;|\alpha_{k,e}^{r}\rangle\;
= \;\sin^{2}\theta\;|V_{k}|^{2}\;
\eqno(4.24)$$
$$\frac{1}{{\cal N}_{k}}\langle\alpha_{k,e}^{r}(t)|\;N_{\alpha_{e}}^{k,r}\;
|\alpha_{k,e}^{r}(t)\rangle\;=\;1
-\;R_{k}\;\sin^{2}2\theta
\;\sin^{2}\left(\frac{\Delta\omega}{2}t\right)\; \eqno(4.25)$$
$$\frac{1}{{\cal N}_{k}}\langle\alpha_{k,e}^{r}(t)|\;N_{\alpha_{\mu}}^{k,r}\;
|\alpha_{k,e}^{r}(t)\rangle\;=
\;R_{k}\;\sin^{2}2\theta
\;\sin^{2}\left(\frac{\Delta\omega}{2}t\right) + \frac{1}{{\cal N}_{k}}
\langle\alpha_{k,e}^{r}|\;N_{\alpha_{\mu}}^{k,r}\;
|\alpha_{k,e}^{r}\rangle\; \eqno(4.26)$$
$$\frac{1}{{\cal N}_{k}}\langle\alpha_{k,e}^{r}(t)|\;
N_{\alpha_{e}}^{k,r}\;|\alpha_{k,e}^{r}(t)\rangle\; +
\;\frac{1}{{\cal N}_{k}}\langle\alpha_{k,e}^{r}(t)|\;N_{\alpha_{\mu}}^{k,r}
\;|\alpha_{k,e}^{r}(t)\rangle\;-
\;\frac{1}{{\cal N}_{k}}\langle\alpha_{k,e}^{r}|\;N_{\alpha_{\mu}}^{k,r}\;
|\alpha_{k,e}^{r}\rangle=1. \eqno(4.27)$$
respectively, with $R_{k}\equiv\frac{|U_{k}|^{2}}{{\cal N}_{k}}=
\frac{1-|V_{k}|^{2}}{1-\sin^{2}\theta\;|V_{k}|^{2}}$.

In conclusion, eqs.(4.19) and (4.20) exhibit the corrections to the
flavor oscillations
coming from the condensate contributions.

The conventional (approximate) results (4.7) and (4.8) are
obtained when the condensate
contributions are missing (in the $|V_{k}| \rightarrow 0$ limit).

Notice that the fraction of $\alpha_{e}^{k,r}$ particles which is
condensed into the vacuum $|0\rangle_{1,2}$ is "frozen", i.e. does not
oscillate in time, as it is easily seen by noting that
$e^{-iH_{1,2} t} |0\rangle_{1,2} = |0\rangle_{1,2} $ (cf. also eq. (4.15)).

It is remarkable that the corrections depend on the modulus $k$ through
$|U_{k}|^{2} =1-\frac {1}{2} Z_{k}$.
Since $Z_{k} \rightarrow 0$ when $k \rightarrow \infty$,
these corrections disappear in the infinite momentum limit. However,
for finite $k$, the oscillation amplitude is depressed by a factor
$|U_{k}|^{2}$: the  depression factor ranges
from $1$ to $\frac{1}{2}$ depending on $k$ and on the masses values,
according to the behaviour of the
$Z_{k}$ function. It is an interesting question to ask if an experimental
test may show such a dependence of the flavor oscillation
amplitude.

We stress that the limit $k \rightarrow \infty$ of eqs.(4.19) and (4.20) gives
an exact result and is not the result of the finite volume approximation.

Since the correction factor is related to the vacuum condensate,
we see that the vacuum acts as a "momentum (or spectrum) analyzer" for the
oscillating neutrinos: neutrinos with $k\gg\sqrt{m_{1}m_{2}}$ oscillate
more than neutrinos with $k\simeq\sqrt{m_{1}m_{2}}$,
due to the vacuum structure. Such a
vacuum spectral analysis effect may sum up to other effects (such as MSW
effect [8] in the matter) in depressing or enhancing neutrino oscillations.

The above scheme is easily generalized to the oscillations
in the matter. As well known [1,2,8],
the two flavors oscillations picture is modified
due to the different interaction of $\nu_{e}$ and $\nu_{\mu}$
with the electrons of the medium. There are two contributions to this
interaction: the first one, coming from neutral current, amounts to
$-G_{F} n_{n}/\sqrt{2}$ and it is present for both $\nu_{e}$ and $\nu_{\mu}$;
the second one, coming from charged current, is given by $\sqrt{2}G_{F} n_{e}$
and it is present only for $\nu_{e}$. Here $n_{e}$ and $n_{n}$ are the
electron and neutron densities, respectively.
This produces a difference in the effective
masses of $\nu_{e}$ and $\nu_{\mu}$, which can be expressed in terms of new
free fields $\tilde{\nu}_{1}$ and $\tilde{\nu}_{2}$ with new masses
$\tilde{m}_{1}$ and $\tilde{m}_{2}$ and a new mixing angle $\tilde{\theta}$.

The mixing relations (2.1) are thus rewritten as:
$$\nu_{e}(x) = {\tilde\nu}_{1}(x) \; \cos{\tilde\theta}
+ {\tilde\nu}_{2}(x) \; \sin{\tilde\theta}  $$
$$\nu_{\mu}(x) =- {\tilde\nu}_{1}(x) \; \sin{\tilde\theta}   +
{\tilde\nu}_{2}(x)\; \cos{\tilde\theta}\;.  \eqno(4.28)$$

The tilde quantities are calculated by diagonalizating the hamiltonian
of $\nu_{e}$ and $\nu_{\mu}$ in matter, which is:
$$H_{e,\mu}^{matter}=H_{e,\mu}^{vacuum} -G_{F} n_{n}/\sqrt{2}
\left(\begin{array}{cc} 1 & 0 \\ 0 & 1\end{array}\right)
+ \sqrt{2}G_{F} n_{e}
\left(\begin{array}{cc} 1 & 0 \\ 0 & 0\end{array}\right)\;. \eqno(4.29)$$

The matter and the vacuum parameters are related as follows [1,2,8]:

$$ \sin^{2}2{\tilde\theta}\simeq \sin^{2}2\theta
\left( \frac{\Delta m^{2}}{\Delta {\tilde m}^{2}} \right) \eqno(4.30)$$

$$\Delta {\tilde m}^{2} = \left[ (D- \Delta m^{2} \cos 2\theta)^{2} +
(\Delta m^{2} \sin 2\theta)^{2}\right]^{\frac{1}{2}} \eqno(4.31)$$
and
$${\tilde m}^{2}_{1,2}= \frac{1}{2}\left(m_{1}^{2}+m_{2}^{2} +
D \mp \Delta {\tilde m}^{2} \right) \eqno(4.32)$$
where $D= 2 \sqrt{2} G_{F} n_{e} k$.

When $D= 2 \sqrt{2} G_{F} n^{crit}_{e} k = \Delta m^{2} \cos2\theta $
there is resonance and $\sin^{2}2{\tilde\theta} $ goes to unity
(MSW effect).

In our scheme it is possible to treat oscillations in matter starting with
mixing relations (4.28) and repeating all the procedure described above. So we
can use all the results obtained in our treatment simply
by substituting $\theta$, $m_{1}$,
$m_{2}$ with the corresponding tilde quantities. In particular, the oscillation
formula becomes in the matter:
$$\langle\alpha_{k,e}^{r}(t)|\;N_{\alpha_{e}}^{k,r}\;
|\alpha_{k,e}^{r}(t)\rangle\;=\left(\;1 -\;\sin^{2}{\tilde\theta}\;
|{\tilde V }_{k}|^{2}\right)
-\;|{\tilde U }_{k}|^{2}\;
\sin^{2}2{\tilde \theta}
\;\sin^{2}\left(\frac{\Delta{\tilde\omega}}{2}t\right)\;. \eqno(4.33)$$

\bigskip
\bigskip

{\bf 5  Three flavors fermion mixing}

\medskip

The extension of our discussion to three flavors is complicated by the
proliferation of terms in the explicit computation of the
quantities of interest. However, it is possible to extract some
results from the structure of the annihilators, without explicitly
writing the expression for the vacuum state.

Among the various possible parameterizations of the three fields mixing matrix,
we choose to work with the following one:
$$ M=
\left(\begin{array}{ccc}
 c_{12}c_{13}& s_{12}c_{13} & s_{13}e^{i\delta}  \\
-s_{12}c_{23}-c_{12}s_{23}s_{13}e^{i\delta} &
c_{12}c_{23}-s_{12}s_{23}s_{13}e^{i\delta} & s_{23}c_{13} \\
s_{12}s_{23}-c_{12}c_{23}s_{13}e^{i\delta} &
-c_{12}s_{23}-s_{12}c_{23}s_{13}e^{i\delta} & c_{23}c_{13}
\end{array}\right)\eqno(5.1)$$
with $c_{ij}\equiv\cos\theta_{ij}$, $s_{ij}\equiv\sin\theta_{ij}$,
since it is the familiar parameterization of CKM matrix [1].

To generate the M matrix, we define
$$ G_{12}(\theta_{12}) = exp(\theta_{12} L_{12})\;\;\;,\;\;\;
  G_{23}(\theta_{23}) = exp(\theta_{23} L_{23})\;\;\;,\;\;\;
  G_{13}(\theta_{13}) = exp(\theta_{13} L_{13})\eqno(5.2)$$
where

$$  L_{12}\equiv \int d^{3}x  \left(\nu_{1}^{\dag}(x)
\nu_{2}(x) - \nu_{2}^{\dag}(x) \nu_{1}(x)
\right)\eqno(5.3a)$$
$$  L_{23}\equiv \int d^{3}x  \left(\nu_{2}^{\dag}(x)
\nu_{3}(x) - \nu_{3}^{\dag}(x) \nu_{2}(x)
\right)\eqno(5.3b)$$
$$  L_{13}\equiv \int d^{3}x  \left(e^{i\delta}\;\nu_{1}^{\dag}(x)
\nu_{3}(x) - e^{-i\delta}\;
\nu_{3}^{\dag}(x) \nu_{1}(x)\right) \eqno(5.3c)$$
so that
$$ \nu^{\alpha}_{e}(x)=
G_{12}^{-1}G_{13}^{-1}G_{23}^{-1}\;
\nu^{\alpha}_{1}(x)\;
G_{23}G_{13}G_{12}\eqno(5.4a)$$
$$ \nu^{\alpha}_{\mu}(x)=
G_{12}^{-1}G_{13}^{-1}G_{23}^{-1}\;
\nu^{\alpha}_{2}(x)
\;G_{23}G_{13}G_{12}\eqno(5.4b)$$
$$ \nu^{\alpha}_{\tau}(x)=
G_{12}^{-1}G_{13}^{-1}G_{23}^{-1}\;
\nu^{\alpha}_{3}(x)\;
G_{23}G_{13}G_{12}\;.\eqno(5.4c)$$

The matrix M is indeed obtained by using the following relations:
$$\left[\nu^{\alpha}_{1}(x) , L_{12}\right]=\nu^{\alpha}_{2}(x) \;\;\;,\;\;\;
\left[\nu^{\alpha}_{1}(x) , L_{23}\right]=0                    \;\;\;,\;\;\;
\left[\nu^{\alpha}_{1}(x) , L_{13}\right]=e^{i\delta}\;\nu^{\alpha}_{3}(x)
\eqno(5.5a)$$
$$\left[\nu^{\alpha}_{2}(x) , L_{12}\right]=-\nu^{\alpha}_{1}(x) \;\;\;,\;\;\;
\left[\nu^{\alpha}_{2}(x) , L_{23}\right]= \nu^{\alpha}_{3}(x)  \;\;\;,\;\;\;
\left[\nu^{\alpha}_{2}(x) , L_{13}\right]=0 \eqno(5.5b)$$
$$\left[\nu^{\alpha}_{3}(x) , L_{12}\right]=0                    \;\;\;,\;\;\;
\left[\nu^{\alpha}_{3}(x) , L_{23}\right]=-\nu^{\alpha}_{2}(x)  \;\;\;,\;\;\;
\left[\nu^{\alpha}_{3}(x) , L_{13}\right]=
- e^{-i\delta}\;\nu^{\alpha}_{1}(x)\;. \eqno(5.5c)$$

Notice that the phase $\delta$ is unavoidable for three fields mixing, while
it can be incorporated in the fields definition for two fields mixing.

The vacuum in the flavor representation is:
$$ |0\rangle_{e\mu\tau}=G_{12}^{-1}G_{13}^{-1}G_{23}^{-1}\;|0\rangle_{123}\;.
\eqno(5.6)$$

We do not give here the explicit form of this state, which is very complicated
and is a combination of all possible couples $\alpha^{r\dag}_{k,i}
\beta^{r\dag}_{-k,j}$ with $i,j=1,2,3$. Nevertheless, we can obtain physical
informations from the structure of the annihilators $\alpha^{r}_{k,l}$,
$\beta^{r}_{k,l}$
($l=e, \mu, \tau$). In the reference frame $k=(0,0,|k|)$ we obtain
(see Appendix E):
$$\alpha_{k,e}^{r}=c_{12}c_{13}\;\alpha_{k,1}^{r}
+ s_{12}c_{13}\left(U^{k*}_{12}\;\alpha_{k,2}^{r} +\epsilon^{r}
V^{k}_{12}\;\beta_{-k,2}^{r\dag}\right)
+ e^{i\delta}\;s_{13}\left(U^{k*}_{13}\;\alpha_{k,3}^{r} +\epsilon^{r}
V^{k}_{13}\;\beta_{-k,3}^{r\dag}\right)\;,\eqno(5.7a)$$

$$\alpha_{k,\mu}^{r}=\left(c_{12}c_{23}- e^{-i\delta}
\;s_{12}s_{23}s_{13}\right)\;\alpha_{k,2}^{r}
- \left(s_{12}c_{23}+e^{-i\delta}\;c_{12}s_{23}s_{13}\right)
\left(U^{k}_{12}\;\alpha_{k,1}^{r}
-\epsilon^{r} V^{k}_{12}\;\beta_{-k,1}^{r\dag}\right)
+$$
$$+ \;s_{23}c_{13}\left(U^{k*}_{23}\;\alpha_{k,3}^{r} +
\epsilon^{r} V^{k}_{23}\;\beta_{-k,3}^{r\dag}\right)\;,
\eqno(5.7b)$$

$$\alpha_{k,\tau}^{r}=c_{23}c_{13}\;\alpha_{k,3}^{r}
- \left(c_{12}s_{23}+e^{-i\delta}\;s_{12}c_{23}s_{13}\right)
\left(U^{k}_{23}\;\alpha_{k,2}^{r}
-\epsilon^{r} V^{k}_{23}\;\beta_{-k,2}^{r\dag}\right)+$$
$$+\;\left(s_{12}s_{23}- e^{-i\delta}\;c_{12}c_{23}s_{13}\right)
\left(U^{k}_{13}\;\alpha_{k,1}^{r}
-\epsilon^{r} V^{k}_{13}\;\beta_{-k,1}^{r\dag}\right)\;,
\eqno(5.7c)$$

$$\beta^{r}_{-k,e}=c_{12}c_{13}\;\beta_{-k,1}^{r}
+ s_{12}c_{13}\left(U^{k*}_{12}\;\beta_{-k,2}^{r}
-\epsilon^{r}V^{k}_{12}\;\alpha_{k,2}^{r\dag}\right)
+e^{-i\delta}\; s_{13}\left(U^{k*}_{13}\;\beta_{-k,3}^{r}
-\epsilon^{r} V_{13}^{k}\;\alpha_{k,3}^{r\dag}\right)\;,
\eqno(5.7d)$$

$$\beta^{r}_{-k,\mu}=\left(c_{12}c_{23}- e^{i\delta}\;
s_{12}s_{23}s_{13}\right)\;\beta_{-k,2}^{r}
- \left(s_{12}c_{23}+e^{i\delta}\;c_{12}s_{23}s_{13}\right)
\left(U^{k}_{12}\;\beta_{-k,1}^{r} +\epsilon^{r}\;
V^{k}_{12}\;\alpha_{k,1}^{r\dag}\right) +$$
$$+\; s_{23}c_{13}\left(U^{k*}_{23}\;\beta_{-k,3}^{r} -
\epsilon^{r}\; V^{k}_{23}\;\alpha_{k,3}^{r\dag}\right)\;,
\eqno(5.7e)$$

$$\beta^{r}_{-k,\tau}=c_{23}c_{13}\;\beta_{-k,3}^{r}
- \left(c_{12}s_{23}+e^{i\delta}\;s_{12}c_{23}s_{13}\right)
\left(U^{k}_{23}\;\beta_{-k,2}^{r} +
\epsilon^{r} V^{k}_{23}\;\alpha_{k,2}^{r\dag}\right)+$$
$$+\;\left(s_{12}s_{23}- e^{i\delta}\;c_{12}c_{23}s_{13}\right)
\left(U^{k}_{13}\;\beta_{-k,1}^{r} +
\epsilon^{r} V^{k}_{13}\;\alpha_{k,1}^{r\dag}\right)\;.\eqno(5.7f)$$
\smallskip

{}From eqs.(5.7) we observe that, in contrast with the case of two flavors
mixing,
the condensation
densities are now different for different flavors (cf. eq.(4.17)):

$$\;\;\;\;\;\;\;\;\;\;\;\;\;_{123}\langle0|N^{k,r}_{\alpha_{e}}|0\rangle_{123}=
\;_{123}\langle0|N^{k,r}_{\beta_{e}}|0\rangle_{123}=
 s^{2}_{12}c^{2}_{13}\;|V^{k}_{12}|^{2}+ s^{2}_{13}\;|V^{k}_{13}|^{2}\;,
 \eqno(5.8a)$$

\pagebreak
$$\;_{123}\langle0|N^{k,r}_{\alpha_{\mu}}|0\rangle_{123}=
\;_{123}\langle0|N^{k,r}_{\beta_{\mu}}|0\rangle_{123}=$$
$$=\left|s_{12}c_{23}+e^{-i\delta}\;c_{12}s_{23}s_{13}\right|^{2}
\:|V^{k}_{12}|^{2}
+ s^{2}_{23}c^{2}_{13}\;|V^{k}_{23}|^{2}\;, \eqno(5.8b)$$

$$\;_{123}\langle0|N^{k,r}_{\alpha_{\tau}}|0\rangle_{123}=
\;_{123}\langle0|N^{k,r}_{\beta_{\tau}}|0\rangle_{123}=$$
$$=\left|c_{12}s_{23}+e^{-i\delta}\;s_{12}c_{23}s_{13}\right|^{2}
|V^{k}_{23}|^{2}
+ \left|s_{12}s_{23}- e^{-i\delta}\;c_{12}c_{23}s_{13}\right|^{2}
|V^{k}_{13}|^{2}\;. \eqno(5.8c)$$

\medskip

To study the three flavors neutrino oscillations, we observe that, in the
finite volume approximation
$$|\alpha_{k,e}^{r}\rangle\;=\;
c_{12}c_{13}\; |\alpha_{k,1}^{r}\rangle +
s_{12}c_{13}\;|\alpha_{k,2}^{r}\rangle +e^{i\delta}\;s_{13}\;|\alpha_{k,3}^{r}
\rangle\;. \eqno(5.9)$$
The time evolution of this state is given by
$$|\alpha_{k,e}^{r}(t)\rangle\;=\;
c_{12}c_{13}\; e^{-i\omega_{1} t} |\alpha_{k,1}^{r}\rangle +
s_{12}c_{13}\;e^{-i\omega_{2} t} |\alpha_{k,2}^{r}\rangle +
e^{i\delta}\;s_{13}\;e^{-i\omega_{3} t} |\alpha_{k,3}^{r}\rangle\;.
\eqno(5.10)$$
The number operator is
$$N_{\alpha_{e}}^{k,r}=\alpha_{k,e}^{r\dag}\alpha_{k,e}^{r}= $$
$$=c^{2}_{12}c^{2}_{13}   \;\alpha_{k,1}^{r\dag}\alpha_{k,1}^{r} +
c_{12}s_{12}c^{2}_{13} \;\alpha_{k,1}^{r\dag}\alpha_{k,2}^{r} +
e^{i\delta}\;c_{12}c_{13}s_{13}     \;\alpha_{k,1}^{r\dag}\alpha_{k,3}^{r} +$$
$$+c_{12}s_{12}c^{2}_{13} \;\alpha_{k,2}^{r\dag}\alpha_{k,1}^{r} +
s^{2}_{12}c^{2}_{13}    \;\alpha_{k,2}^{r\dag}\alpha_{k,2}^{r} +
e^{i\delta}\;s_{12}c_{13}s_{13}      \;\alpha_{k,2}^{r\dag}\alpha_{k,3}^{r} +$$
$$+e^{-i\delta}\;c_{12}c_{13}s_{13} \;\alpha_{k,3}^{r\dag}\alpha_{k,1}^{r} +
e^{-i\delta}\;s_{12}c_{13}s_{13}  \;\alpha_{k,3}^{r\dag}\alpha_{k,2}^{r} +
s^{2}_{13}                 \;\alpha_{k,3}^{r\dag}\alpha_{k,3}^{r}\;.
\eqno(5.11)$$
and thus
$$\langle\alpha_{k,e}^{r}(t)|\;N_{\alpha_{e}}^{k,r}\;
|\alpha_{k,e}^{r}(t)\rangle\;=\;
c^{4}_{12}c^{4}_{13}+s^{4}_{12}c^{4}_{13}+s^{4}_{13}+
c^{2}_{12}s^{2}_{12}c^{4}_{13}
(e^{-i(\omega_{2}-\omega_{1}) t}+e^{+i(\omega_{2}-\omega_{1}) t})+$$
$$+c^{2}_{12}c^{2}_{13}s^{2}_{13}
(e^{-i(\omega_{3}-\omega_{1}) t}+e^{+i(\omega_{3}-\omega_{1}) t})
+s^{2}_{12}c^{2}_{13}s^{2}_{13}
(e^{-i(\omega_{3}-\omega_{2}) t}+e^{+i(\omega_{3}-\omega_{2}) t})$$
i.e.
$$\langle\alpha_{k,e}^{r}(t)|\;N_{\alpha_{e}}^{k,r}\;
|\alpha_{k,e}^{r}(t)\rangle\;=\;
1 - \cos^{4}\theta_{13}\;\sin^{2}2\theta_{12}\;
\sin^{2}\left(\frac{\omega_{2}-\omega_{1}}{2}t\right) +$$
$$-\cos^{2}\theta_{12}\;\sin^{2}2\theta_{13}\;
\sin^{2}\left(\frac{\omega_{3}-\omega_{1}}{2}t\right)
-\sin^{2}\theta_{12}\;\sin^{2}2\theta_{13}\;
\sin^{2}\left(\frac{\omega_{3}-\omega_{2}}{2}t\right)\;. \eqno(5.12)$$

On the other hand, the QFT computations give

$$\langle\alpha_{k,e}^{r}|\;N_{\alpha_{e}}^{k,r}\;
|\alpha_{k,e}^{r}\rangle\;=\;1- s^{2}_{12}c_{13}^{2}\;|V^{k}_{12}|^{2}
-s^{2}_{13}\;|V^{k}_{13}|^{2}\;, \eqno(5.13)$$
$$\;_{123}\langle0|\;N_{\alpha_{e}}^{k,r}\;
|0\rangle_{123}\;=\; s^{2}_{12}c_{13}^{2}\;|V^{k}_{12}|^{2}
+s^{2}_{13}\;|V^{k}_{13}|^{2}\;, \eqno(5.14)$$
and

$$\langle\alpha_{k,e}^{r}(t)|\;N_{\alpha_{e}}^{k,r}\;
|\alpha_{k,e}^{r}(t)\rangle\;=\;
c^{4}_{12}c^{4}_{13}+
|U^{k}_{12}|^{2}\;s^{4}_{12}c^{4}_{13}+
|U^{k}_{13}|^{2}\;s^{4}_{13}+
|V^{k}_{12}|^{2}\;c^{2}_{12}s^{2}_{12}c^{4}_{13}+
|V^{k}_{13}|^{2}\;c^{2}_{12}c^{2}_{13}s^{2}_{13}+$$
$$+\left(|U^{k}_{12}|^{2}\;|V^{k}_{13}|^{2}\;+\;|U^{k}_{13}|^{2}
\;|V^{k}_{12}|^{2}\;
\right)s^{2}_{12}c^{2}_{13}s^{2}_{13}
+|U^{k}_{12}|^{2}\;c^{2}_{12}s^{2}_{12}c^{4}_{13}
(e^{-i(\omega_{2}-\omega_{1}) t}+e^{+i(\omega_{2}-\omega_{1}) t})+$$
$$+|U^{k}_{13}|^{2}\;c^{2}_{12}c^{2}_{13}s^{2}_{13}
(e^{-i(\omega_{3}-\omega_{1}) t}+e^{+i(\omega_{3}-\omega_{1}) t})
+|U^{k}_{12}|^{2}\;|U^{k}_{13}|^{2}\;s^{2}_{12}c^{2}_{13}s^{2}_{13}
(e^{-i(\omega_{3}-\omega_{2}) t}+e^{+i(\omega_{3}-\omega_{2}) t}),$$
namely
$$\langle\alpha_{k,e}^{r}(t)|\;N_{\alpha_{e}}^{k,r}\;
|\alpha_{k,e}^{r}(t)\rangle\;=\;\left(1- \sin^{2}\theta_{12}
\cos^{2}\theta_{13}\;|V^{k}_{12}|^{2}
-\sin^{2}\theta_{13}\;|V^{k}_{13}|^{2}\right)+$$
$$
-|U^{k}_{12}|^{2}\; \cos^{4}\theta_{13}\;\sin^{2}2\theta_{12}\;
\sin^{2}\left(\frac{\omega_{2}-\omega_{1}}{2}t\right)
-|U^{k}_{13}|^{2}\; \cos^{2}\theta_{12}\;\sin^{2}2\theta_{13}\;
\sin^{2}\left(\frac{\omega_{3}-\omega_{1}}{2}t\right)+$$
$$
-|U^{k}_{12}|^{2}\;|U^{k}_{13}|^{2}\;
\sin^{2}\theta_{12}\;\sin^{2}2\theta_{13}\;
\sin^{2}\left(\frac{\omega_{3}-\omega_{2}}{2}t\right)\;, \eqno(5.15)$$
which is to be compared with eq.(5.12).


\bigskip

{\bf 6  Conclusions}

\medskip

In this paper we have studied the fermion mixing transformations in the QFT
framework. In particular we have considered the Pontecorvo
mixing transformations
for neutrino Dirac fields [5]. In the LSZ formalism of QFT [3,6,7] the
Fock space of definite flavor states is shown to be unitarily
inequivalent in the
infinite volume limit to the Fock space of definite mass states. The flavor
states are obtained as condensate of massive neutrino-antineutrino
pairs and exhibit the structure of $SU(2)$ generalized coherent states [4].
The condensation density is computed as a function of the mixing angle, of the
momentum modulus and of the neutrino masses.

The neutrino oscillation formula is derived and its amplitude turns out to be
momentum dependent. We suggest that such a result may be object of
experimental investigation.

In the $k \rightarrow \infty$ limit the momentum dependence disappears and the
oscillation formula reproduces the usual one. Notice however that the
oscillation formula we obtain in the limit $k \rightarrow \infty$ is exact and
is not the result of the finite volume approximation used in the conventional
treatment.

Since the oscillation term is related in our
analysis to the vacuum condensate, the vacuum acts as a "momentum (or spectrum)
analyzer" for the oscillating neutrinos. Such a vacuum spectral analysis
effect may contribute in depressing or enhancing neutrino oscillations.

We observe that the functional dependence of the oscillating term on the
momentum is such that, if experimentally tested, may give indication on the
neutrino masses since the function $Z_{k}=2\left(1-|U_{k}|^{2}\right)$
(cf. eqs.(3.10) and (4.19)) has a maximum at
$k= \sqrt{m_{1}m_{2}}$.

Although the  physically relevant quantities are given by expectation values
of the observables, nevertheless it is interesting to observe that the ratio
of the amplitudes of the $|\alpha_{k,1}^{r}\rangle$ and
$|\alpha_{k,2}^{r}\rangle$ components of the state
$|\alpha_{k,e}^{r}(t)\rangle$
is constant in time, as can be seen from eq.(4.14), and that such a feature
persists even in the limit $k \rightarrow \infty$ (i.e.
$|U_{k}| \rightarrow 1$) where, however, the oscillation formula (4.19) reduces
to the usual one (4.7). This is in contrast with the finite volume
approximation where "decoherence" between the components
$|\alpha_{k,1}^{r}\rangle$ and $|\alpha_{k,2}^{r}\rangle$ arises from the phase
factor $exp(-i\Delta\omega \;t)$ (see eq.(4.5)).

We have
shown that our discussion can be extended to the oscillations in matter and the
corresponding oscillation formula is obtained.

We have also studied the three flavors mixing and
have obtained the
corresponding oscillation formula, which also in this case is momentum
dependent.

We stress the crucial r\^ole played by the existence in QFT of
infinitely many unitarily inequivalent representations of the canonical
anti-commutation rules. We have in fact explicitly shown
that the neutrino mixing
transformations map state spaces which are unitarily inequivalent in the
infinite volume limit. In this way we realize the limit of validity of the
identification of the vacuum state for definite mass neutrinos with
the vacuum state for definite flavor neutrinos. We point out that such
an identification is actually an approximation since the flavor vacuum has
the structure of an $SU(2)$ generalized coherent state and
it is only allowed at finite volume. The vacua identification is
meaningless in QFT since the mass and the flavor space are
unitarily inequivalent.

Finally, we observe that although our discussion has been focused on the
neutrino mixing, nevertheless it can be extended, with due changes, to other
fermion mixing transformations, such as the CKM mixing transformations. In this
last case, the so called free fields in the LSZ formalism are to be understood
as the asymptotically free quark fields.

The study of the QFT for mixing of boson fields with different masses is also
in progress. In that case preliminary results show [9] that relations
analogous to eqs.(3.2) and (3.6) hold so that we have a non trivial vacuum
structure. In the boson case we find $|U_{k}|=\cosh\sigma_{k}$ and
$|V_{k}|=\sinh\sigma_{k}$, with $\sigma_{k}= \frac{1}{2} \log
\left(\frac{\omega_{k,1}}{\omega_{k,2}}\right)$ where $\omega_{k,i}$,
$i=1,2$, is the boson energy.

\bigskip

{\bf Acknowledgements}

\medskip

We are grateful to Professors S. M. Bilenky and A. Perelomov for useful
discussions. We thank E. Alfinito and A. Iorio for their comments and
their help in some computations.

\newpage

{\bf Appendix A}

\medskip

Using the algebra (2.19) and the relations (2.27), we have:
$$\begin{array}{ll}
S_{+}^{k}S_{-}^{k}|0\rangle_{1,2} = S_{-}^{k}S_{+}^{k}|0\rangle_{1,2}
&(S_{+}^{k})^{2}(S_{-}^{k})^{2}|0\rangle_{1,2} =
(S_{-}^{k})^{2}(S_{+}^{k})^{2}|0\rangle_{1,2} \\ \\
(S_{+}^{k})^{2}S_{-}^{k}|0\rangle_{1,2} = S_{-}^{k}(S_{+}^{k})^{2}
|0\rangle_{1,2} +2S_{+}^{k}|0\rangle_{1,2}\\ \\
(S_{-}^{k})^{2}S_{+}^{k}|0\rangle_{1,2} = S_{+}^{k}(S_{-}^{k})^{2}
|0\rangle_{1,2} +2S_{-}^{k}|0\rangle_{1,2}
\\ \\
(S_{+}^{k})^{3}S_{-}^{k}|0\rangle_{1,2} = 6(S_{+}^{k})^{2}|0\rangle_{1,2}
& (S_{+}^{k})^{4}S_{-}^{k}|0\rangle_{1,2} =0\\ \\
(S_{+}^{k})^{3}(S_{-}^{k})^{2}|0\rangle_{1,2} =
6S_{-}^{k}(S_{+}^{k})^{2}|0\rangle_{1,2}
&(S_{-}^{k})^{3}(S_{+}^{k})^{2}|0\rangle_{1,2} =
6S_{+}^{k}(S_{-}^{k})^{2}|0\rangle_{1,2}\\ \\
(S_{+}^{k})^{4}(S_{-}^{k})^{2}|0\rangle_{1,2} =24(S_{+}^{k})^{2}|0\rangle_{1,2}
&(S_{+}^{k})^{5}(S_{-}^{k})^{2}|0\rangle_{1,2} =0\\ \\
S_{+}^{k}S_{-}^{k}(S_{+}^{k})^{2}|0\rangle_{1,2}=
4(S_{+}^{k})^{2}|0\rangle_{1,2}
&S_{-}^{k}S_{+}^{k}(S_{-}^{k})^{2}|0\rangle_{1,2}=
4(S_{-}^{k})^{2}|0\rangle_{1,2}\\ \\
S_{3}^{k}S_{-}^{k}S_{+}^{k}|0\rangle_{1,2}=0
&S_{3}^{k}(S_{+}^{k})^{2}(S_{-}^{k})^{2}|0\rangle_{1,2} = 0\\ \\
(S_{3}^{k})^{n}S_{-}^{k}|0\rangle_{1,2} = (-1)^{n} S_{-}^{k}|0\rangle_{1,2}
&(S_{3}^{k})^{n}(S_{-}^{k})^{2}|0\rangle_{1,2} =
(-2)^{n} (S_{-}^{k})^{2}|0\rangle_{1,2}\\ \\
(S_{3}^{k})^{n}S_{+}^{k}|0\rangle_{1,2} = S_{+}^{k}|0\rangle_{1,2}
&(S_{3}^{k})^{n}(S_{+}^{k})^{2}|0\rangle_{1,2} =
2^{n} (S_{+}^{k})^{2}|0\rangle_{1,2}\\ \\
S_{3}^{k}S_{-}^{k}(S_{+}^{k})^{2}|0\rangle_{1,2}=
S_{-}^{k}(S_{+}^{k})^{2}|0\rangle_{1,2}
&S_{3}^{k}S_{+}^{k}(S_{-}^{k})^{2}|0\rangle_{1,2}=
-S_{+}^{k}(S_{-}^{k})^{2}|0\rangle_{1,2}
\end{array}$$
Use of the above relations gives eq.(2.28).


\bigskip
\medskip

{\bf Appendix B}

\medskip

Wave functions and $Z_{k}$:

$$u_{k,i}^{r}(t)=\hat{u}_{k,i}^{r} e^{-i\omega_{k,i} \; t}=A_{i}
\left(\begin{array}{c}
 \xi^{r} \\ \\
\frac{\bar{\sigma}\cdot\bar{k}}{\omega_{k,i}+m_{i}} \xi^{r}
\end{array}\right)e^{-i\omega_{k,i} \; t}\;\;\;,\;\;\;
v_{k,i}^{r}(t)=\hat{v}_{k,i}^{r} e^{i\omega_{k,i} \; t}=A_{i}
\left(\begin{array}{c}
\frac{\bar{\sigma}\cdot\bar{k}}{\omega_{k,i}+m_{i}} \xi^{r} \\ \\
\xi^{r}
\end{array}\right)e^{i\omega_{k,i} \; t}$$
$$\xi_{1}=\left(\begin{array}{c} 1 \\ 0 \end{array}\right)\;\;,\;\;
\xi_{2}=\left(\begin{array}{c} 0 \\ 1 \end{array}\right)\;\;\;,\;\;\;
A_{i}\equiv
\left(\frac{\omega_{k,i}+m_{i}}{2\omega_{k,i}}\right)^{\frac{1}{2}}
\;\;\;,\;\;\; i=1,2\;,\;\; r=1,2\;.$$
$$\hat{v}_{-k,1}^{1\dag} \hat{u}_{k,2}^{1}=
- \hat{v}_{-k,1}^{2\dag} \hat{u}_{k,2}^{2}=
A_{1}A_{2}
\left(\frac{-k_{3}}{\omega_{k,1}+m_{1}} +\frac{k_{3}}{\omega_{k,2}+m_{2}}
\right)$$
$$\hat{v}_{-k,1}^{1\dag} \hat{u}_{k,2}^{2}=
\left(\hat{v}_{-k,1}^{2\dag} \hat{u}_{k,2}^{1}\right)^{*}
=A_{1}A_{2}
\left(\frac{-k_{1}+ik_{2}}{\omega_{k,1}+m_{1}}
+\frac{k_{1}-ik_{2}}{\omega_{k,2}+m_{2}}\right)$$

Using the above relations eq.(2.32) is obtained.

Eq.(2.31) follows if one observe that
$$ (S_{-}^{k})^{2} |0\rangle_{1,2}
=2\left[
\left(u^{1\dag}_{k,2}v^{2}_{-k,1}\right)
\left(u^{2\dag}_{k,2}v^{1}_{-k,1}\right)
-\left(u^{1\dag}_{k,2}v^{1}_{-k,1}\right)
\left(u^{2\dag}_{k,2}v^{2}_{-k,1}\right) \right]
\alpha^{1\dag}_{k,2}\beta^{2\dag}_{-k,1}
\alpha^{2\dag}_{k,2}\beta^{1\dag}_{-k,1} |0\rangle_{1,2}=$$
$$=Z_{k}\;e^{2i(\omega_{k,1}+\omega_{k,2} ) t}\;
\alpha^{1\dag}_{k,2}\beta^{2\dag}_{-k,1}
\alpha^{2\dag}_{k,2}\beta^{1\dag}_{-k,1}  |0\rangle_{1,2}\;.$$

\bigskip
\medskip

{\bf Appendix C}

\medskip

Relationships for the number operator and eq.(3.2):

$$\begin{array}{ll}
\left[N^{k}_{1},S_{+}^{k}\right]=\sum_{r,s}\left(
(u^{r \dag}_{k,1}u^{s}_{k,2})\alpha^{r\dag}_{k,1}\alpha^{s}_{k,2}+
(u^{r \dag}_{k,1}v^{s}_{-k,2})\alpha^{r\dag}_{k,1}\beta^{s \dag}_{-k,2}
\right) \\ \\
\left[N^{k}_{1},S_{-}^{k}\right]=-\sum_{r,s}\left(
(u^{r \dag}_{k,2}u^{s}_{k,1})\alpha^{r\dag}_{k,2}\alpha^{s}_{k,1}+
(v^{r \dag}_{-k,2}u^{s}_{k,1})\beta^{r}_{-k,2}\alpha^{s}_{k,1}\right) \\ \\
\left[N^{k}_{1},S_{-}^{k}\right]|0\rangle_{1,2}=
N^{k}_{1}S_{-}^{k}|0\rangle_{1,2} =N^{k}_{1}(S_{-}^{k})^{2}|0\rangle_{1,2} =0
\\ \\
\left[N^{k}_{1},S_{-}^{k}\right]S_{-}^{k}|0\rangle_{1,2}=
\left[N^{k}_{1},S_{-}^{k}\right](S_{-}^{k})^{2}|0\rangle_{1,2}=
\left[N^{k}_{1},S_{+}^{k}\right](S_{+}^{k})^{2}|0\rangle_{1,2}=0
\;\;\;\;\;\;\;\;\;\;\;\;\;\\ \\
\left[N^{k}_{1},S_{+}^{k}\right]|0\rangle_{1,2} =S_{+}^{k}|0\rangle_{1,2}
\;\;\;\;,\;\;\;\;
\left[N^{k}_{1},S_{+}^{k}\right]S_{+}^{k}|0\rangle_{1,2}=
(S_{+}^{k})^{2}|0\rangle_{1,2}\\ \\
N^{k}_{1}S_{+}^{k}|0\rangle_{1,2} =S_{+}^{k}|0\rangle_{1,2}
\;\;\;\;\;\;\;,\;\;\;\;\;
N^{k}_{1}(S_{+}^{k})^{2}|0\rangle_{1,2}=2(S_{+}^{k})^{2}|0\rangle_{1,2}
\;\;\;\;\;\;\;\;\;\;\;\;\;\;\;\;\;
\end{array}$$

\medskip

$$\begin{array}{l}
\;_{1,2}\langle0|S_{-}^{k}N^{k}_{1}S_{+}^{k}|0\rangle_{1,2}=
 Z_{k} \\ \\
\;_{1,2}\langle0|S_{-}^{k}N^{k}_{1}S_{-}^{k}(S_{+}^{k})^{2}|0\rangle_{1,2}=
\;_{1,2}\langle0|(S_{-}^{k})^{2}S_{+}^{k}N^{k}_{1}S_{+}^{k}|0\rangle_{1,2}=
(Z_{k})^{2}\\ \\
\;_{1,2}\langle0|S_{+}^{k}S_{-}^{k}N^{k}_{1}S_{+}^{k}S_{-}^{k}
|0\rangle_{1,2}=Z_{k}\;+\;\frac{1}{2}(Z_{k})^{2}\\ \\
\;_{1,2}\langle0|(S_{-}^{k})^{2}N^{k}_{1}(S_{+}^{k})^{2}|0\rangle_{1,2}=
2\;(Z_{k})^{2}\\ \\
\;_{1,2}\langle0|(S_{-}^{k})^{2}S_{+}^{k}N^{k}_{1}
S_{-}^{k}(S_{+}^{k})^{2}|0\rangle_{1,2}=6\; (Z_{k})^{2}\\ \\
\;_{1,2}\langle0|(S_{+}^{k})^{2}S_{-}^{k}N^{k}_{1}
S_{+}^{k}(S_{-}^{k})^{2}|0\rangle_{1,2}=2\;(Z_{k})^{2}\\ \\
\;_{1,2}\langle0|(S_{+}^{k})^{2}(S_{-}^{k})^{2}N^{k}_{1}
S_{+}^{k}S_{-}^{k}|0\rangle_{1,2}=
\;_{1,2}\langle0|S_{+}^{k}S_{-}^{k}N^{k}_{1}(S_{+}^{k})^{2}(S_{-}^{k})^{2}
|0\rangle_{1,2}=
4\;(Z_{k})^{2}\\ \\
\;_{1,2}\langle0|(S_{+}^{k})^{2}(S_{-}^{k})^{2}N^{k}_{1}
(S_{+}^{k})^{2}(S_{-}^{k})^{2}|0\rangle_{1,2}=
24\;(Z_{k})^{2}\;.
\end{array}$$

Eq.(3.2) is then obtained as
$$\begin{array}{l}
_{e,\mu}\langle0|N^{k}_{\alpha_{1}}|0\rangle_{e,\mu}=
Z_{k}(\sin^{2}\theta\cos^{2}\theta + \sin^{4}\theta)+\\
+(Z_{k})^{2}\sin^{4}\theta(-\cos^{2}\theta + 2\sin^{2}\theta\cos^{2}\theta
-2\sin^{2}\theta +\frac{1}{2}\cos^{4}\theta+\frac{1}{2}
+\frac{3}{2}\sin^{4}\theta )=\\
=Z_{k}\sin^{2}\theta \;.
\end{array}$$

\bigskip
\medskip

{\bf Appendix D}

\medskip

For the calculation  of $|0\rangle_{e,\mu}^{k}$ it is useful to
choose $k=(0,0,|k|)$. In this reference frame
the operators $S_{+}^{k}$, $S_{-}^{k}$, $S_{3}^{k}$ are written as follows:
$$S_{+}^{k}\equiv\sum_{k,r}S_{+}^{k,r}=$$
$$=\sum_{r} \left( U^{*}_{k}\; \alpha^{r\dag}_{k,1} \alpha^{r}_{k,2}
- \epsilon^{r}\; V^{*}_{k}\; \beta^{r}_{-k,1}  \alpha^{r}_{k,2}
+ \epsilon^{r}\; V_{k}\;\alpha^{r\dag}_{k,1}\beta^{r\dag}_{-k,2}
+ U_{k}\; \beta^{r}_{-k,1}\beta^{r\dag}_{-k,2}\right)$$
$$S_{-}^{k}\equiv\sum_{k,r}S_{-}^{k,r}=$$
$$=\sum_{r} \left( U_{k}\; \alpha^{r\dag}_{k,2} \alpha^{r}_{k,1}
+ \epsilon^{r}\; V^{*}_{k}\; \beta^{r}_{-k,2}  \alpha^{r}_{k,1}
- \epsilon^{r}\; V_{k}\;\alpha^{r\dag}_{k,2}\beta^{r\dag}_{-k,1}
+ U_{k}^{*}\; \beta^{r}_{-k,2}\beta^{r\dag}_{-k,1}\right)$$
$$S_{3}\equiv \sum_{k,r} S_{3}^{k,r}
=\frac{1}{2}\sum_{k,r}\left(\alpha^{r\dag}_{k,1}\alpha^{r}_{k,1}
-\beta^{r\dag}_{-k,1}\beta^{r}_{-k,1}
-\alpha^{r\dag}_{k,2}\alpha^{r}_{k,2} + \beta^{r\dag}_{-k,2}\beta^{r}_{-k,2}
\right)\;, $$
where $U_{k}$, $V_{k}$ have been defined in eqs.(3.7)-(3.10) and
$\epsilon^{r}=(-1)^{r}$. It is easy to show that the $su(2)$ algebra
holds for $S_{\pm}^{k,r}$ and $S_{3}^{k,r}$,
which means that the $su_{k}(2)$ algebra given in eqs.(2.19)
splits into $r$ disjoint $su_{k,r}(2)$ algebras.
Using the Gaussian decomposition, $|0\rangle_{e,\mu}^{k} $ can be written as
$$|0\rangle_{e,\mu}^{k} = \prod_{r} exp(-tan\theta \; S_{+}^{k,r})
exp(-2 ln \: cos\theta \; S_{3}^{k,r})
\;exp(tan\theta \; S_{-}^{k,r})|0\rangle_{1,2} $$
where $0\leq \theta < \frac{\pi}{2}$.
The final expression for $|0\rangle_{e,\mu}^{k}$ in terms of
$S^{k,r}_{\pm}$ and $S^{k,r}_{3}$ is then
$$|0\rangle_{e,\mu}^{k}=
\prod_{r}\left[ 1 + \sin\theta \cos\theta
\left(S_{-}^{k,r} - S_{+}^{k,r}\right)
-\sin^{2}\theta \; S_{+}^{k,r}S_{-}^{k,r}\right]|0\rangle_{1,2}\;,$$
from which we finally obtain eq.(3.11).


{\bf Appendix E}

\smallskip

Useful relations for three flavors mixing.

We work in the frame $k=(0,0,|k|)$ and for simplicity
we omit the $k$ and the helicity indices.

$\left\{\begin{array}{l}
G_{23}^{-1} \alpha_{1} G_{23}=\alpha_{1} \\ \\
G_{13}^{-1} \alpha_{1} G_{13}=
 	c_{13}\;\alpha_{1}\;+e^{i\delta}\;s_{13}\;
\left( U_{13}^{*}\; \alpha_{3}\;
+\epsilon_{r}\;V_{13}\; \beta^{\dag}_{3}\right) \\ \\
G_{12}^{-1} \alpha_{1} G_{12}=
 	c_{12}\;\alpha_{1}\;+\;s_{12}\;
\left( U_{12}^{*}\; \alpha_{2}\;+\epsilon_{r}\;V_{12}\; \beta^{\dag}_{2}\right)
\end{array} \right.$

\smallskip

$\left\{\begin{array}{l}
G_{23}^{-1} \alpha_{2} G_{23}=
	c_{23}\;\alpha_{2}\;+\;s_{23}\;
\left( U_{23}^{*}\; \alpha_{3}\;
+\epsilon_{r}\;V_{23}\; \beta^{\dag}_{3}\right) \\ \\
G_{13}^{-1} \alpha_{2} G_{13}=\alpha_{2} \\ \\
G_{12}^{-1} \alpha_{2} G_{12}=
 	c_{12}\;\alpha_{2}\;-\;s_{12}\;
\left( U_{12}\; \alpha_{1}\;-\epsilon_{r}\;V_{12}\; \beta^{\dag}_{1}\right)
\end{array}\right.$

\smallskip

$\left\{\begin{array}{l}
G_{23}^{-1} \alpha_{3} G_{23}=
	c_{23}\;\alpha_{3}\;-\;s_{23}\;
\left( U_{12}\; \alpha_{2}\;
-\epsilon_{r}\;V_{12}\; \beta^{\dag}_{2}\right) \\ \\
G_{13}^{-1} \alpha_{3} G_{13}=
 	c_{13}\;\alpha_{3}\;-e^{-i\delta}\;s_{13}\;
\left( U_{13}\; \alpha_{1}\;-\epsilon_{r}\;V_{13}\; \beta^{\dag}_{1}\right)
\\ \\
G_{12}^{-1} \alpha_{3} G_{12}=\alpha_{3}
\end{array}\right.$

\smallskip

$\left\{\begin{array}{l}
G_{23}^{-1} \beta^{\dag}_{1} G_{23}=\beta^{\dag}_{1} \\ \\
G_{13}^{-1} \beta^{\dag}_{1} G_{13}=
 	c_{13}\;\beta^{\dag}_{1}\;+e^{i\delta}\;s_{13}\;
\left( U_{13}\; \beta^{\dag}_{3}\;-\epsilon_{r}\;V_{13}^{*}\;
\alpha_{3}\right) \\ \\
G_{12}^{-1} \beta^{\dag}_{1} G_{12}=
 	c_{12}\;\beta^{\dag}_{1}\;+\;s_{12}\;
\left( U_{12}\; \beta^{\dag}_{2}\;-\epsilon_{r}\;V_{12}^{*}\; \alpha_{2}\right)
\end{array}\right.$

\smallskip

$\left\{\begin{array}{l}
G_{23}^{-1} \beta^{\dag}_{2} G_{23}=
	c_{23}\;\beta^{\dag}_{2}\;+\;s_{23}\;
\left( U_{23}\; \beta^{\dag}_{3}\;
-\epsilon_{r}\;V_{23}^{*}\; \alpha_{3}\right)  \\ \\
G_{13}^{-1} \beta^{\dag}_{2} G_{13}=\beta^{\dag}_{2} \\ \\
G_{12}^{-1} \beta^{\dag}_{2} G_{12}=
 	c_{12}\;\beta^{\dag}_{2}\;-\;s_{12}\;
\left( U_{12}^{*}\; \beta^{\dag}_{1}\;
+\epsilon_{r}\;V_{12}^{*}\; \alpha_{1}\right)
\end{array}\right.$

\smallskip

$\left\{\begin{array}{l}
G_{23}^{-1} \beta^{\dag}_{3} G_{23}=
	c_{23}\;\beta^{\dag}_{3}\;-\;s_{23}\;
\left( U_{23}^{*}\; \beta^{\dag}_{2}\;
+\epsilon_{r}\;V_{23}^{*}\; \alpha_{2}\right) \\ \\
G_{13}^{-1} \beta^{\dag}_{3} G_{13}=
 	c_{13}\;\beta^{\dag}_{3}\;-e^{-i\delta}\;s_{13}\;
\left( U_{13}^{*}\; \beta^{\dag}_{1}\;
+\epsilon_{r}\;V_{13}^{*}\; \alpha_{1}\right) \\ \\
G_{12}^{-1} \beta^{\dag}_{3} G_{12}=\beta^{\dag}_{3}
\end{array}\right.$

$$ $$
with
$$V^{k}_{ij}=|V^{k}_{ij}|\;e^{i(\omega_{k,j}+\omega_{k,i})t}\;\;\;\;,\;\;\;\;
U^{k}_{ij}=|U^{k}_{ij}|\;e^{i(\omega_{k,j}-\omega_{k,i})t}$$
$$|U^{k}_{ij}|=\left(\frac{\omega_{k,i}+m_{i}}{2\omega_{k,i}}\right)
^{\frac{1}{2}}
\left(\frac{\omega_{k,j}+m_{j}}{2\omega_{k,j}}\right)^{\frac{1}{2}}
\left(1+\frac{k^{2}}{(\omega_{k,i}+m_{i})(\omega_{k,j}+m_{j})}\right)$$
$$|V^{k}_{ij}|=\left(\frac{\omega_{k,i}+m_{i}}{2\omega_{k,i}}\right)
^{\frac{1}{2}}
\left(\frac{\omega_{k,j}+m_{j}}{2\omega_{k,j}}\right)^{\frac{1}{2}}
\left(\frac{k}{(\omega_{k,j}+m_{j})}-\frac{k}{(\omega_{k,i}+m_{i})}\right)$$
where $i,j=1,2,3$ and $j>i$, and

$$|U^{k}_{ij}|^{2}+|V^{k}_{ij}|^{2}=1 \;\;\;\;\;,\;\;\;\; i=1,2,3 \;\; j>i $$
$$\left(V^{k}_{23}V^{k*}_{13}+U^{k*}_{23}U^{k}_{13}\right)
= U^{k}_{12}\;\;\;\;\;,\;\;\;\;\;
\left(V^{k}_{23}U^{k*}_{13}-U^{k*}_{23}V^{k}_{13}\right)
=- V^{k}_{12}$$
$$\left(U^{k}_{12}U^{k}_{23}-V^{k*}_{12}V^{k}_{23}\right)
= U^{k}_{13}\;\;\;\;\;,\;\;\;\;\;
\left(U^{k}_{23}V^{k}_{12}+U^{k*}_{12}V^{k}_{23}\right)
= V^{k}_{13}$$
$$\left(V^{k*}_{12}V^{k}_{13}+U^{k*}_{12}U^{k}_{13}\right)
= U^{k}_{23}\;\;\;\;\;,\;\;\;\;\;
\left(V^{k}_{12}U^{k}_{13}-U^{k}_{12}V^{k}_{13}\right)
=- V^{k}_{23}\;.$$

\newpage

{\bf References}

\medskip

\begin{enumerate}

\item   T.Cheng and L.Li, {\it Gauge Theory of Elementary Particle Physics},
	Clarendon Press, Oxford, 1989 \\
	R.E.Marshak, {\it Conceptual Foundations of Modern Particle Physics},
	World Scientific, Singapore, 1993

\item  	R.Mohapatra and P.Pal, {\it Massive Neutrinos in Physics and
	Astrophysics}, World Scientific, Singapore, 1991 \\
        J.N.Bahcall, {\it Neutrino Astrophysics},
	Cambridge Univ. Press, Cambridge, 1989

\item	C.Itzykson and J.B.Zuber, {\it Quantum Field Theory},
	McGraw-Hill, New York, 1980

\item	A.Perelomov, {\it Generalized Coherent States and Their
        Applications}, Springer-Verlag, Berlin, 1986

\item	S.M.Bilenky and B.Pontecorvo, {\sl Phys. Rep.} {\bf 41} (1978) 225

\item	N.N.Bogoliubov. A.A.Logunov, A.I.Osak and I.T.Todorov,
	{\it General Principles of Quantum Field Theory},
	Kluwer Academic Publishers, Dordrech, 1990

\item	H.Umezawa, H.Matsumoto and M.Tachiki,
	{\it Thermo Field Dynamics and Condensed States}, North-Holland Publ.
        Co., Amsterdam, 1982 \\
	H.Umezawa,
	{\it Advanced Field Theory: Macro, Micro, and Thermal Physics},
	American Institute of Physics, New York, 1993

\item	L.Wolfenstein, {\it Phys. Rev.}{\bf D17} (1978) \\
	S.P.Mikheyev and A.Y.Smirnov, {\it Nuovo Cimento}
	{\bf 9C} (1986) 17

\item  	M.Blasone and G.Vitiello, {\it Mixing Transformations in Quantum Field
       	Theory}, in preparation, 1994

\end{enumerate}

\end{document}